\documentclass{article}
\usepackage[utf8]{inputenc}
\usepackage{caption}
\usepackage{subcaption}

\usepackage{cite}

\newcommand{\sqrts}{\mbox{$\sqrt{\mathrm{s}}$}}
\newcommand{\axi}{$\overline{\Xi}^+$}
\newcommand{\xim}{$\Xi^-$}
\newcommand{\alam}{$\overline{\Lambda}$}

\newcommand{\lam}{$\Lambda$}
\newcommand{\ks}{$\mathrm{K}^{0}_{\mathrm S}$}

\newcommand{\ppt}{$p_{\rm T}$}

\usepackage{graphicx}

\usepackage{lineno}
\providecommand{\keywords}[1]{\textbf{\textit{Keywords---}} #1}
\usepackage{lineno}

\usepackage{graphicx}
\usepackage{multicol}

\usepackage{lineno}
\begin{document}


\begin{center}
{\Large \bf Model studies of $V^0$ production ratios in $pp$ collisions at \sqrts~= 0.2, 0.9, and 7 TeV}

\vskip1.0cm

M. Ajaz$^{1}${\footnote{Corresponding author: ajaz@awkum.edu.pk; muhammad.ajaz@cern.ch (M. Ajaz)}},
M.U. Ashraf$^{2}$, 
M. Waqas$^{3}${\footnote{Corresponding author: waqas\_phy313@yahoo.com; waqas\_phy313@ucas.ac.cn (M. Waqas)}},
Z. Yasin$^{4}$, 
A.M. Khubrani$^{5}$, 
S. Hassan$^{4}$, 
A. Haj Ismail$^{6,7}$, 
L.L. Li$^{8}$
\\

{\small\it 
$^1$ Department of Physics, Abdul Wali Khan University Mardan, 23200 Mardan, Pakistan,
\\
$^2$ National Center for Physics, Shahdra Valley Road, Islamabad 44000, Pakistan\\

$^3$ School of Nuclear Science and Technology, University of Chinese Academy of Sciences,
Beijing 100049, China\\

$^4$ Pakistan Institute of Nuclear Science and Technology (PINSTECH), Islamabad 44000, Pakistan\\
$^5$ Department of Physics, Faculty of Science, Jazan University, 45142 Jazan, Saudi Arabia\\
$^6$ College of Humanities and Sciences, Ajman University, Ajman 346, UAE\\
$^7$ Nonlinear Dynamic Research Center (NDRC), Ajman University, Ajman 346, UAE\\
$^8$ Department of Basic Sciences, Shanxi Agricultural University, Jinzhong 030801, China
}\\
\end{center}

\vskip1.0cm


\begin{abstract}

This paper has performed a comparative study of $V^0$ ratios between HIJING, Sibyll, and QGSJET model-based event generators. The ratios under study are {\alam}/{\lam}, {\alam}/{\ks} and {\xim}/{\lam} as a function of rapidity $y$, rapidity loss ($\Delta y$) and {\ppt} from $pp$ collisions at \sqrts~= 0.2, 0.9, and 7 TeV and these simulations are then compared with the STAR and LHCb fiducial phase spaces in different {\ppt} regions. Although the models could produce some of the ratios in a limited {\ppt} or $y$ region, none of them completely predicts the experimental results. The QGSJET has good predictions with the data in most of the cases, but since the model does not include the $\Xi$ particle definition, therefore it does not give any predictions for $\Xi$/{\lam} ratios. The extrapolation to the highest possible energies can be studied by re-tuning some basic parameters based on current and previous measurements. These kinds of systematic comparison studies are also helpful in applying certain constraints on the pQCD and non-pQCD-based hadronic event generators to significantly improve the predictions of Standard Model physics at the RHIC and LHC experimental data for the understanding of underlying physics mechanisms in high energy collisions. 
 
\vskip0.5cm
\end{abstract}


\section{Introduction}\label{sec1}
A deconfined, strongly interacting state of matter called Quark-Gluon Plasma (QGP) is conjectured to be produced at high temperatures and densities in collisions recorded at high-energy physics experiments. QGP formation is gauged with several proposed signatures, including production and ratios of strange particles. The transverse momentum ({\ppt}) spectra of strange particles also serve an essential role in the determination of freezeout parameters~\cite{1, 1a, 1b}. Strange hadrons production in relativistic high energy collisions is an invaluable tool to investigate the properties of collision phases since these are not part of the colliding nuclei from the incoming beams. Strange particles have been studied extensively in high energy particle physics as their production and distribution in phase space provide information on the fireball and properties of the created medium~\cite{2}. In the initial stages, after a collision, flavor creation and excitation are mainly responsible for strangeness production at high {\ppt} (gluon splittings dominate in the later evolution). While at low {\ppt}, non-perturbative partonic processes contribute predominantly to the strangeness production~\cite{3, 4}.

High energy proton-proton $(pp)$ collisions provide a simple system to investigate the structure of nuclear matter ~\cite{4a, 4b} and the basic workings of the universe. Many recent developments in fundamental particle physics, including the Higgs boson's discovery, proved the importance of $pp$ collision systems. Additionally, $pp$ collision system serves as a baseline guide for the investigations of complex proton-nucleus $(pA)$ and nucleus-nucleus $(AB)$ collision systems. Transverse momentum ({\ppt}) spectra in $pp$ collisions are used as reference in heavy-ion $(pA, AB)$ collisions for the studies of initial state effects. The high-energy heavy-ion collisions are important to characterize the quark-gluon plasma (QGP). However, many signatures typical to the heavy-ion collisions have also been observed in high multiplicity $pp$ collisions~\cite{1n, 2n} including recent observation of enhanced production of strange and multistrange particles~\cite{3n}. Monte-Carlo's studies of the production of strange and multistrange particles in $pp$ collisions are thus important to characterize not only the $pp$ systems but can further be extended to compare the production of enhanced strangeness in different collision systems. 

Perturbative Quantum-Chromodynamics (pQCD) nicely describes the particle production at high transverse momentum ({\ppt}). In the low ({\ppt}) regions, involving predominant soft processes, phenomenological models are commonly employed. Baryon production in this region notably lacks a full QCD-based description. The ambiguity lies in whether the baryon number should be associated with the valence quarks of a hadron or with its gluonic field. Phenomenological processes involving (anti-)string junctions and hostile C-parity exchanges may give rise to differences between spectra of particles and the corresponding anti-particles. Spectra of (anti-)baryons can provide information regarding competing mechanisms responsible for the baryon production in $pp$ collisions. The anti-baryon to baryon ratio of the hadrons with different valence quark content at different collision energies is considered one of the most direct methods to constrain the baryon production mechanisms~\cite{ALICEn}. Particle ratios simplify the comparison of various experiments performed under different trigger and centrality conditions, and systematic errors associated with absolute particle yields are reduced while working with ratios. Strange to non-strange, mixed particle ratios play an essential role in the quantitative analysis of strangeness production and serve as crucial probes for strangeness enhancement studies~\cite{10}. Particle ratios, such as ($p/\pi$) and ($\Lambda$/$K_S^0$), provide essential insights regarding production mechanisms and the spectral shapes, especially in the intermediate transverse momentum region~\cite{9}. Confronting the model predictions from phenomenological models with experimental observations provide insights on parameters tuning for further improvements in the models~\cite{4}. QGP features and other insights gained from the study of strange particle production ratios may prove to be helpful in the parameter tuning of Monte Carlo models.

In this study anti-baryon to baryon ratios are presented vs the strangeness number: 0$(\bar{p}/p)$, 1$(\bar{\Lambda}/\Lambda)$, 2$(\bar{\Xi}/\Xi)$ and 3$(\bar{\Omega}/\Omega)$. The ratios $(\bar{\Lambda}/\Lambda)$, $(\bar{\Xi}/\Xi)$ as a function of {\ppt} are calculated with hadronic models and contrasted with the corresponding experimental data. The ratio $(\bar{\Lambda}/\Lambda)$ gives the information on the baryon-number transport between proton-proton $pp$ collision state to the final state hadrons. Furthermore, the ratio ($\Lambda$/$K_S^0$) shows the suppression of baryon to meson ratio in strange quark hadronization~\cite{LHCbn}. In this paper, the ratios ($\Lambda$/$K_S^0$) are presented as a function of ({\ppt}), Rapidity $(y)$, and Rapidity loss $(\Delta{y})$ in $pp$ collisions at \sqrts= 200 GeV, 900GeV and 7.0 TeV as calculated with HIJING, Sibyll2.3d and QGSJETII-04 models. A thorough comparison is made between experimental and simulation results and underlying features giving rise to differences in modelling results are discussed.

As proposed, 


Particle production in high energy collisions includes contributions from both the hard processes (explained with perturbative Quantum Chromodynamics, pQCD) and the soft processes that are currently understood with phenomenological models~\cite{9a, 9b}. Confronting predictions from such hadronic models with experimental observations provides insights on parameters tuning for further improvements in the models~\cite{4, 9c, 9d, 9e, 9f, 9g, 9h, 9i, 9j, 9k}. 

The paper is organized into four sections: In section 2, the phenomenological models used for the study are described briefly. Section 3 gives results and discussion, and finally, the results are concluded in section 4.

\section{Methods and Models}
The hadronic models, HIJING, Sibyll2.3d, and QGSJETII-04, used in this study, are briefly described here.

{\bf HIJING} model is a Monte Carlo framework of heavy ions based on the basic structure of C++ and CPU-based parallelism. HIJING is designed to observe the production of jets and multi-particles in $pp$, $pA$,  and $AA$ high energy collisions. HIJING can adjust the parameters for the production of multi-particles in $pp$ collisions to retrieve the essential information of initial conditions in the energy range ($\sqrt{s_{NN}}= 5-2000 $~GeV)~\cite{11}. HIJING, a heavy-ion jet interaction generator, utilizes the LUND model for the fragmentation of jets and the QCD-inspired models for the production of jets. To study the nuclear consequences, the shadowing effect is also incorporated for the structuring of parton functions~\cite{12}. In addition, the model of HIJING includes phenomenology of multi-strings at low $p_{T}$ interaction, thereby combining the physics of fragmentation at inter-mediate energy with perturbative physics at the energy of collisions~\cite{13}. The model also includes the effects of the production of multiple mini jets, soft excitations, and jet interactions in dense hadronic matter~\cite{12, 14}. 

The {\bf Sibyll2.3d} model is based on mini-jet modeling with the ability of multiple hard parton interactions per hadron collision. In this model, the effects of energy-momentum conservation drive the change in the distribution of leading particles. The multiplicity of average particles is high at higher energy~\cite{15} in the Sibyll model. The primary purpose of the event-generator Sibyll2.3d is to take into account the main characteristics of hadronic-particle production and strong interactions. These characteristics are required for the study of air-shower cascades and fluxes of secondary particles, which are produced via the interaction of cosmic rays with the atmosphere of the Earth~\cite{16}. This model consists of a combination of Gribov Regge theory and QCD perturbative field theory with the inclusion of elastic and total cross-sections for p-p, K-p, and $\pi$-p interactions to retrace the new LHC data~\cite{17}. In this version of the Sibyll model, the production of charm-hadrons is also included in the studies of atmospheric neutrinos at high energies. In addition to charm hadrons, the abundance of muons is increased as compared to previous versions to remove well the difference between simulation and data~\cite{16}. Sibyll model also describes the excited states' diffractive production of projectiles and targets, leading-particle distributions with different energies, and particle production in forward phase-space~\cite{17}.

{ \bf QGSJETII-04} is a Monte Carlo event generator based on the Quark-Gluon String model developed for hadronic interactions. The generator of QGSJETII-04 relies on the Quark-Gluon String model, effective field theoretical framework of Gribov-Regge, and LUND algorithm to explain the interactions of high-energy particles and to study phenomena of multiple-scattering. For the nucleus-nucleus interactions and semi-hard processes, the QGSJET model employs the "Semi-hard pomerons" approach~\cite{17, 18}. The coupling of Pomeron in the QGSJETII-04 model is derived from the framework of Gribov Regge theory. The coupling of pomeron-pomeron represents the cascade of parton interaction when pomerons overlap in phase-space. QGSJETTII-04 event generator requires fewer parameters for a simulation~\cite{18} and can incorporate mini jets. The new version of QGJETII-04 is tuned with experimental data from LHC experiments. QGSJETII-04 model reproduces the experimental data at higher-momentum to a good extent, while the particle distributions are over-estimated at lower-momentum~\cite{19, 20}.

\section{Results and discussion}\label{sec3}

The events have been generated for various strange particles, {\ks}, {\lam (\alam)} and {\xim {\axi}} using different models including HIJING, Sibyll and QGSJET at \sqrts~= 0.2, 0.9, and 7 TeV. The pQCD-based model HIJING and cosmic ray air shower based models Sibyll and QGSJET gives different predictions at all energies. In order to validate the MC simulation results, a detailed comparison has been made with that of experimental results from $pp$ collision at \sqrts~= 200 GeV from STAR experiment at RHIC~\cite{21} and LHCb experiment at \sqrts~= 0.9 and 7 TeV~\cite{22}. 

\begin{figure}[ht!]
\centering
\begin{subfigure}[b]{0.49\textwidth}
\centering
\includegraphics[width=\textwidth]{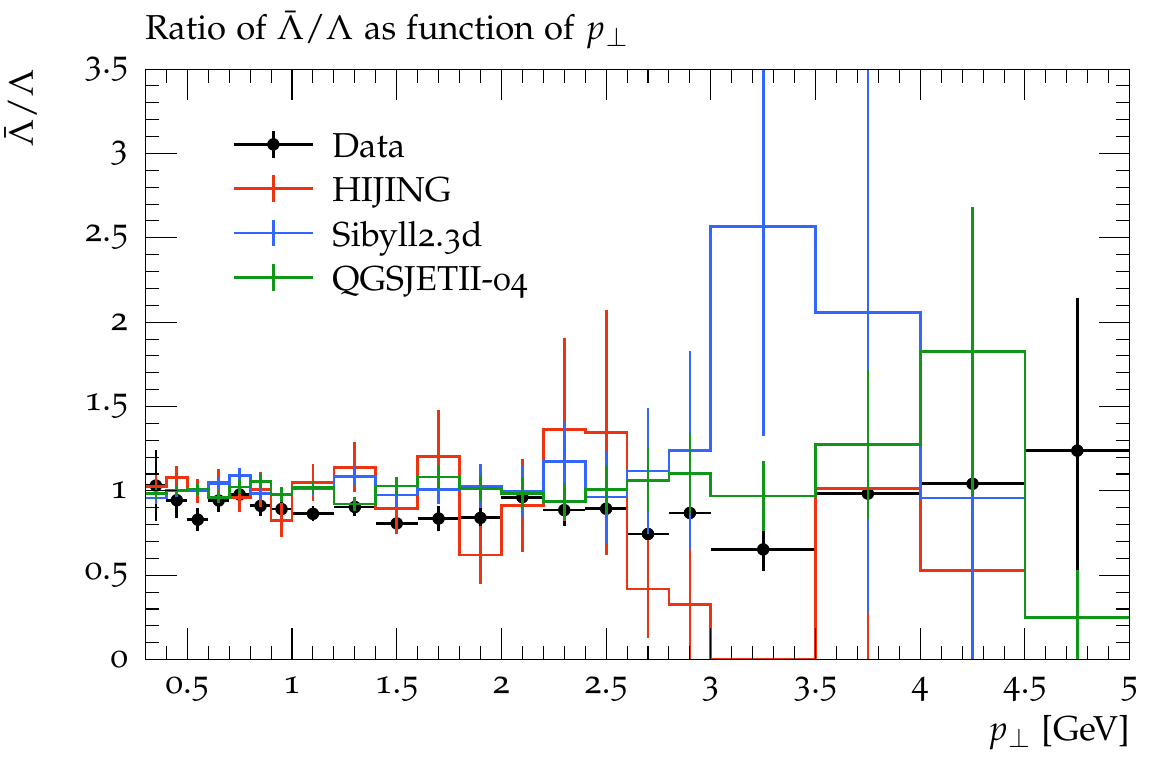}
\caption{{\alam}/{\lam} Ratio as a function of {\ppt}}
\label{fig1a}
\end{subfigure}
\hfill
\begin{subfigure}[b]{0.49\textwidth}
\centering
\includegraphics[width=\textwidth]{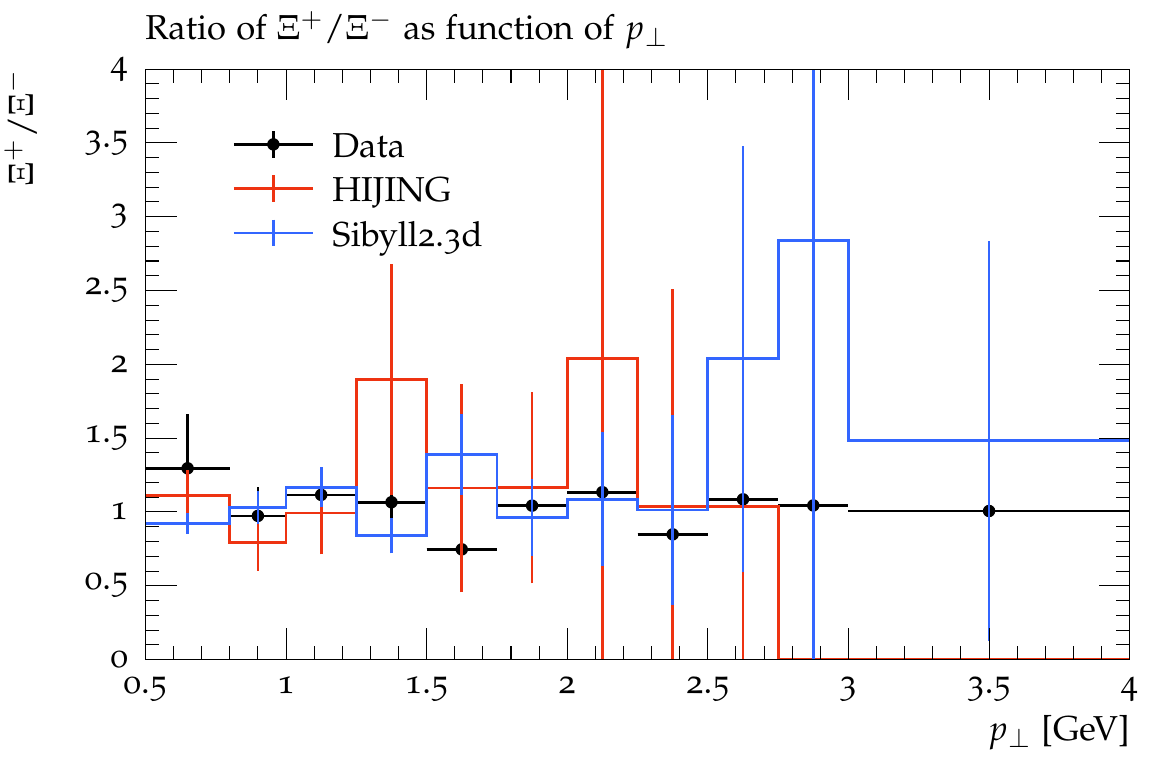}
\caption{{\axi}/{\xim} Ratio as a function of {\ppt}}
\label{fig1b}
\end{subfigure}
\caption{Anti-baryon to baryon ratios as a function of {\ppt} in $pp$ collisions at \sqrts = 200 GeV from STAR experiment in comparison to the model predictions. Black solid markers are the data points and lines of different colors shows different model predictions. }
\label{fig1}
\end{figure}

For a gluon jet, there is no leading baryon against anti-baryon, while in the case of a quark jet, this expectation is reversed as there is a leading baryon against anti-baryon. Therefore, the hadron production mechanisms at high {\ppt} are dominated by the jet fragmentation, and it is a reasonable expectation that with the increase in {\ppt} the $\bar B/ B$ ratio will start to decrease. This decreasing trend in $\bar B/ B$ ratios has been predicted previously~\cite{23}. Figures~\ref{fig1a} and ~\ref{fig1b} shows the anti-baryon to baryon ({\alam}/{\lam}) and {\axi}/{\xim} ratio as a function of {\ppt} respectively. It has been observed that all models describe the experimental data with reasonable agreement while taking into account the error bars and $\chi^2$/$n$ calculated in each case given in Table 1. Among the three, for both the ratios, the Sibyll gives a better description of data, particularly for the {\axi}/{\xim} ratio, than the other model where the QGSJET did not produce any result. However, large error bars in data, as well as models, makes it difficult to observe the ratios follow decreasing trend at high {\ppt}.

\begin{figure}[ht!]
\centering

\includegraphics[width=0.49\textwidth]{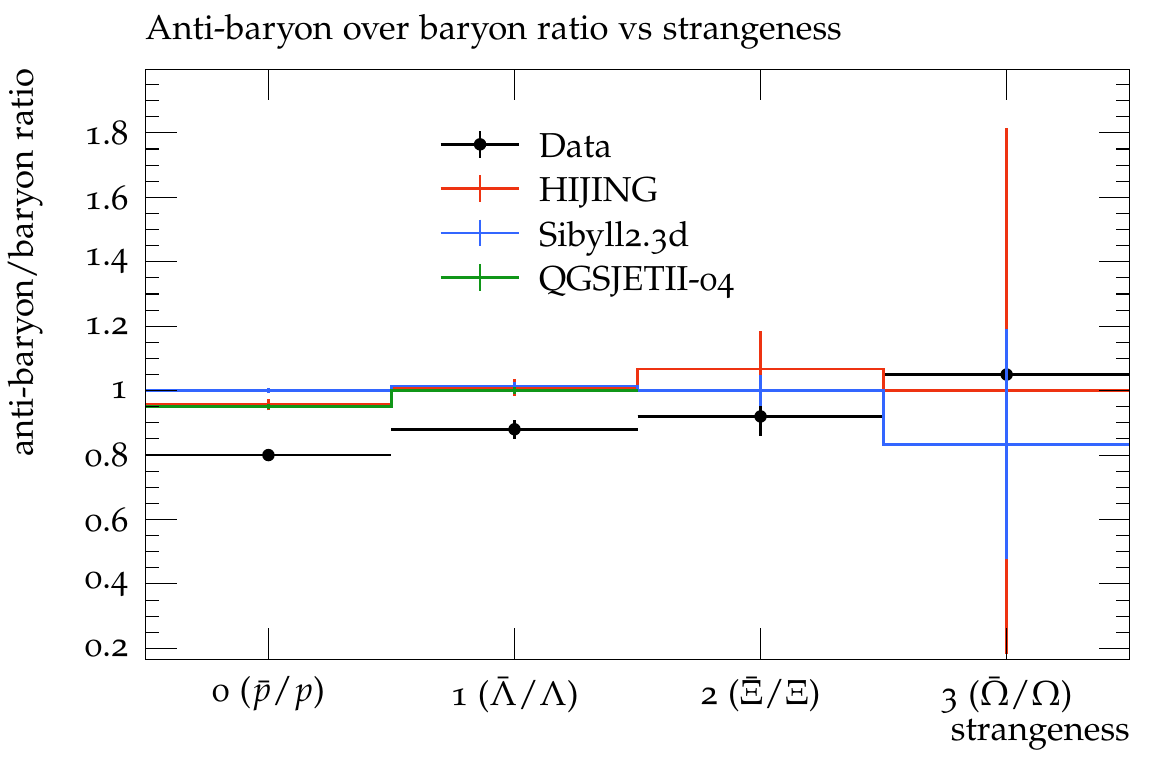}
\caption{Particles ratio and {\ppt} vs mass at \sqrts = 200 GeV}
\label{fig2}
\end{figure}

The mean anti-baron to baryon ($\bar B/ B$) ratio as a function of strangeness content from $pp$ collisions at \sqrts~ = 200 GeV from STAR compared with different models can be seen in figure~\ref{fig2}. The ratio shows a slightly increasing trend with an increase in strangeness content in the case of the experimental data, while the models show a completely different trend. In the case of the experimental data, for protons and {\lam} baryons, the $\bar B/ B$ ratio is not unity, and different parton distribution functions for light quark may explain this deviation~\cite{24}. On the other hand, the ratio from all models does not show a strong dependence on the strange content. This means that the models still need improvements to include the strangeness content. A possible explanation for experimental data as well as the model predictions in fig.~\ref{fig1} and fig.~\ref{fig2} is that the particles are not predominantly produced from the quark-jet fragmentation over the measured {\ppt} range.

In order to understand the model behavior for different particle ratios, a detailed study has been preformed at LHC energies as well. LHCb experiment at the LHC reported the anti-baryon to baryon and baryon to meson ratios as a function of rapidity and {\ppt} in $pp$ collisions at \sqrts~ = 0.9 and 7 TeV~\cite{22}. The {\ppt} is divided into various regions; $0.25 < p_T < 0.65$ GeV/$c$, $0.65 < p_T < 1.0$ GeV/$c$ and $1.0 < p_T < 2.5$ GeV/$c$ in the rapidity $2 < y < 4$ region. The anti-baryon to baryon ({\alam}/{\lam}) and baryon to meson ({\alam}/{\ks}) ratios are then compared with different model predictions at the given {\ppt} and $y$ regions.

\begin{figure}[ht!]
\centering
\includegraphics[width=0.49\textwidth]{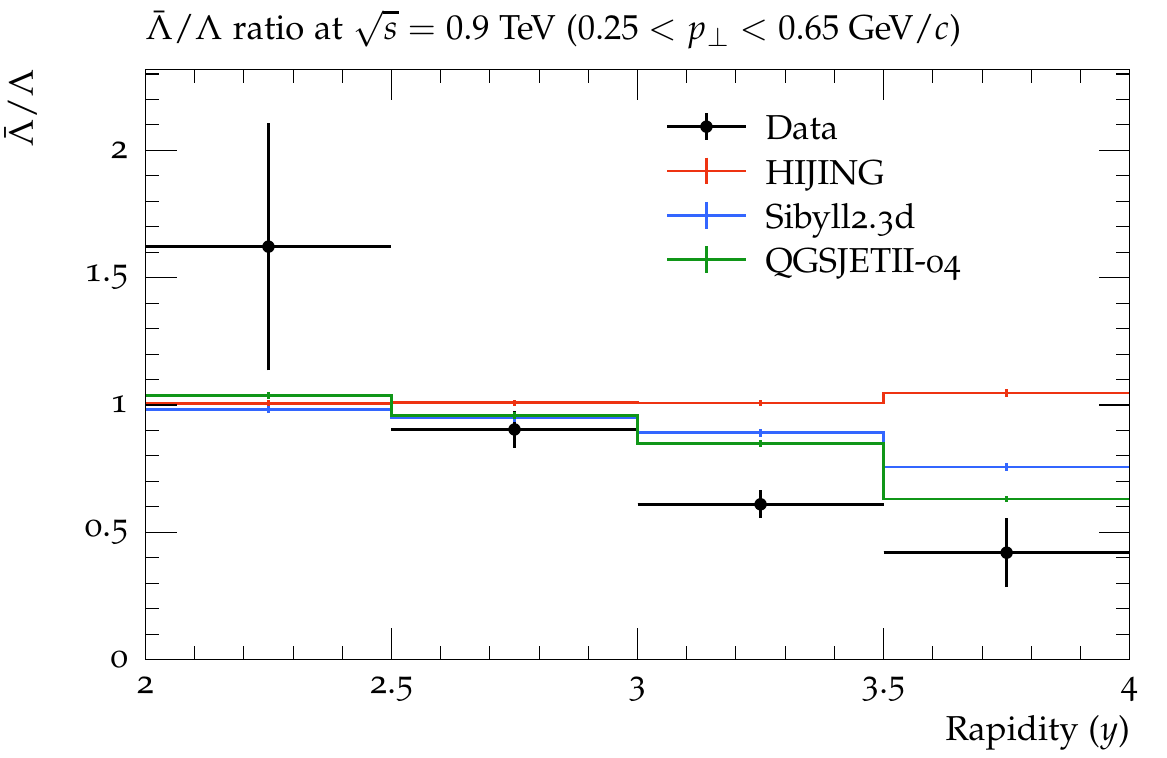}
\includegraphics[width=0.49\textwidth]{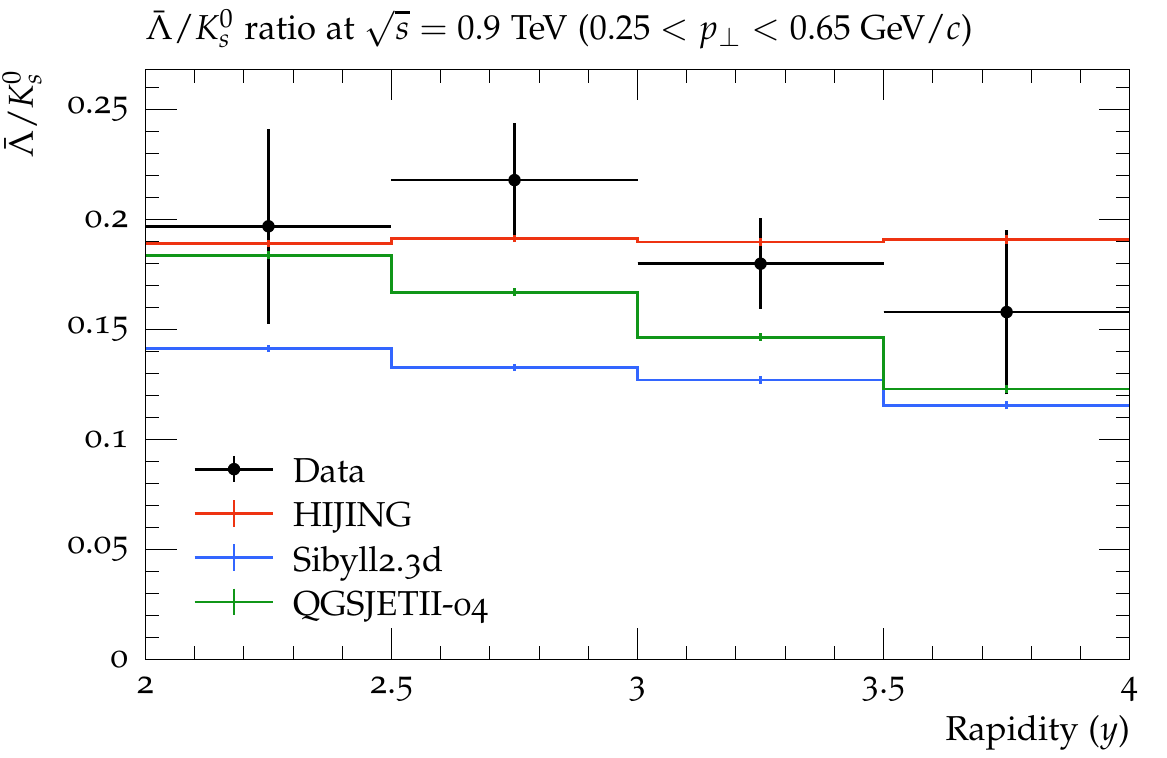}
\includegraphics[width=0.49\textwidth]{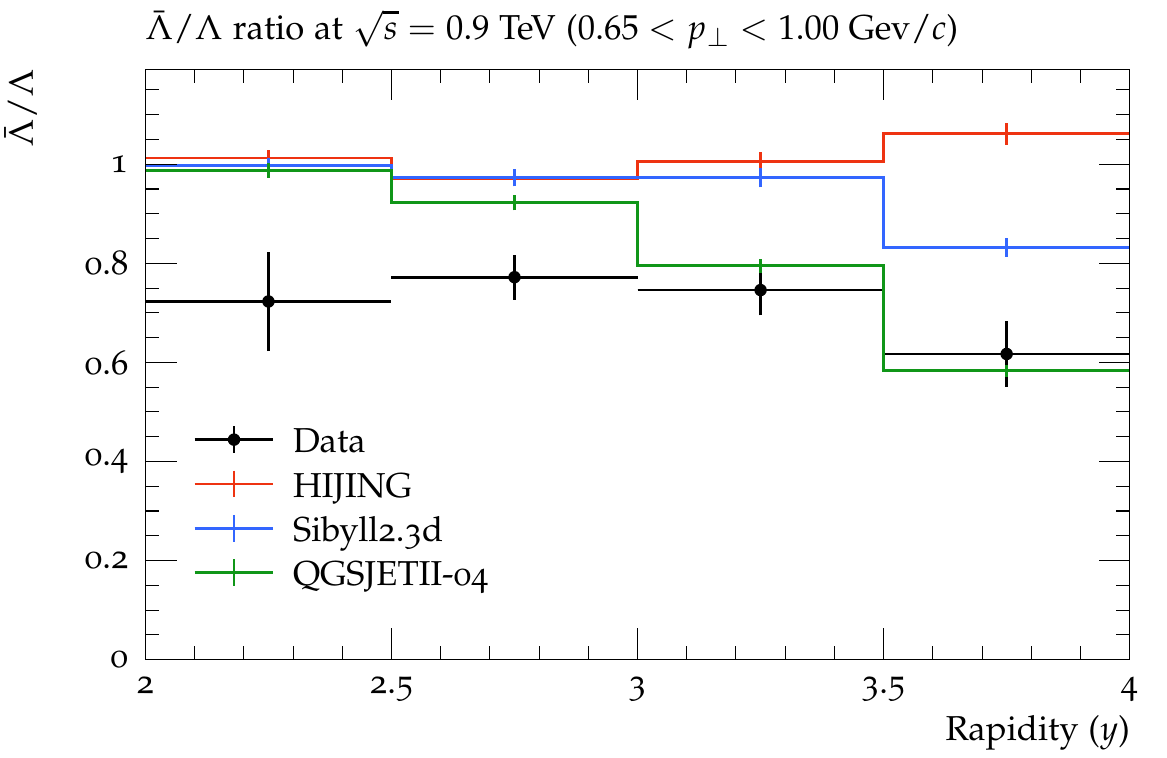}
\includegraphics[width=0.49\textwidth]{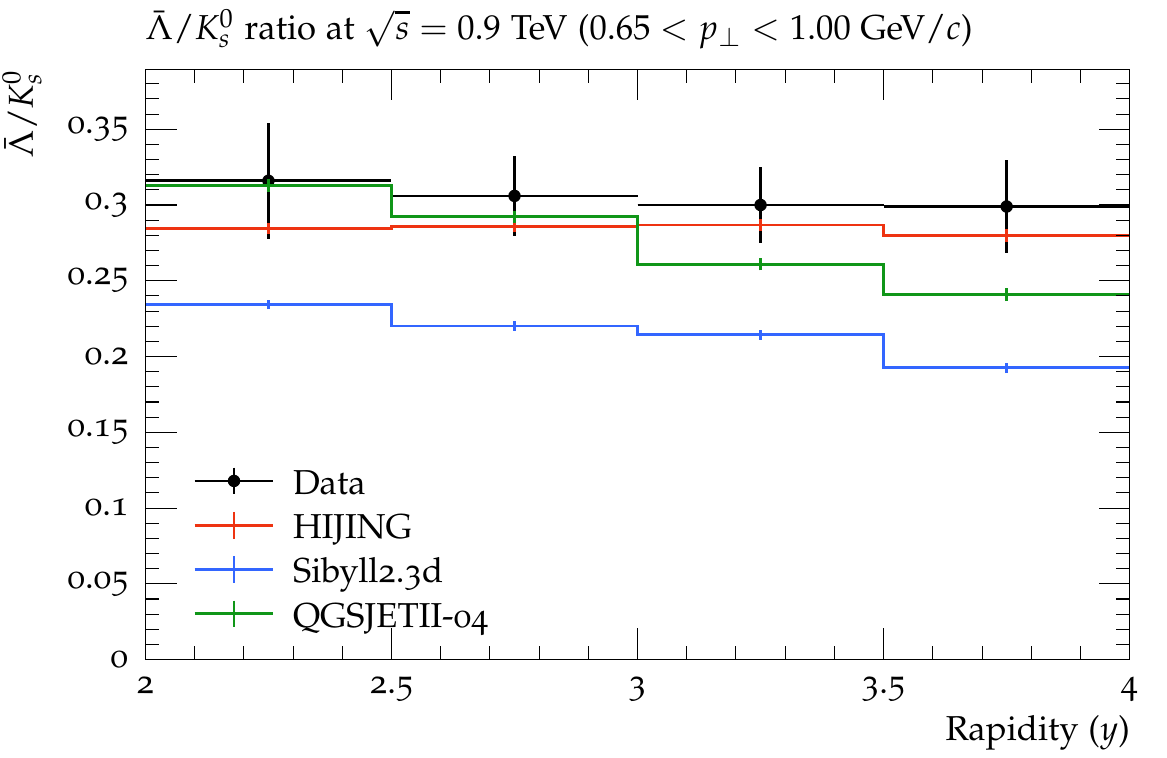}
\includegraphics[width=0.49\textwidth]{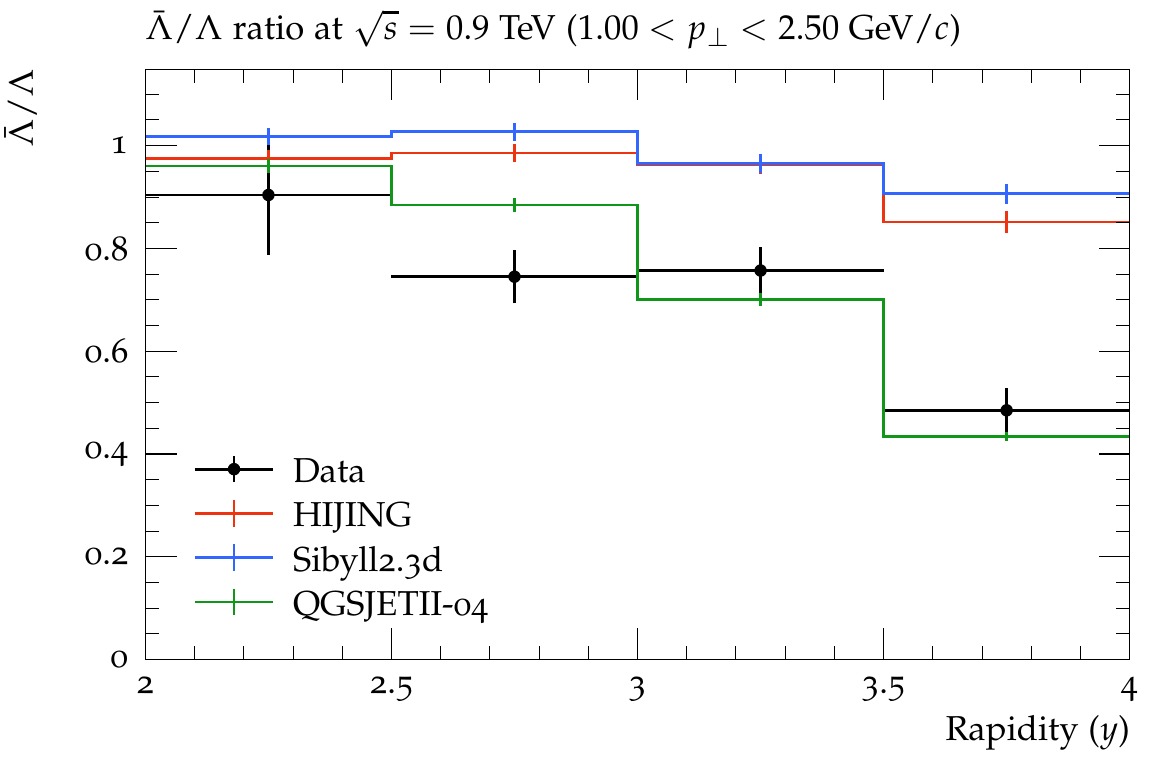}
\includegraphics[width=0.49\textwidth]{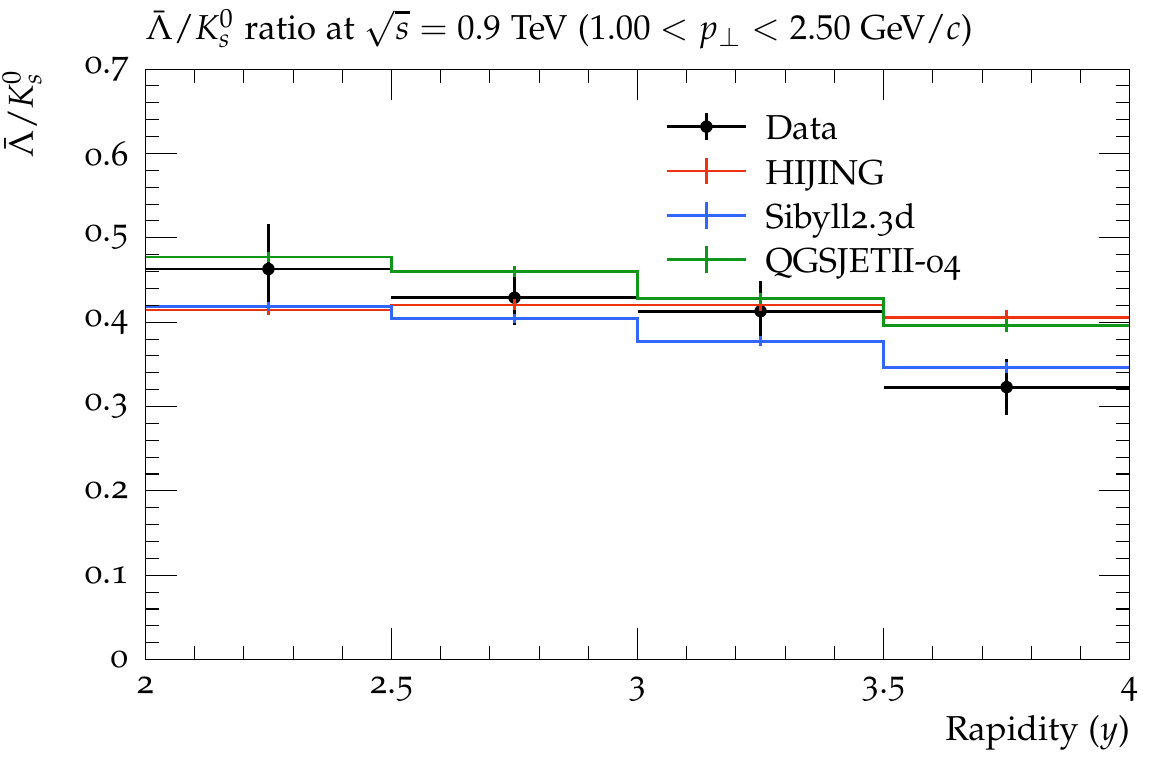}
\caption{(Left column) anti-baryon to baryon ({\alam}/{\lam}) ratios as a function of rapidity ($y$) (Right Column) anti-baryon to meson ({\alam}/{\ks}) ratios in $pp$ collisions \sqrts~ = 0.9 TeV at different {\ppt} regions. Black solid markers are the data points and lines of different colors shows different model predictions.}
\label{fig3}
\end{figure}

Figure~\ref{fig3} (left column) shows the model prediction of anti-baryon to baryon ({\alam}/{\lam}) ratios as a function of rapidity ($y$) in comparison to the data from LHCb experiment~\cite{22}. The $y$ dependence can be seen in the experimental data. It has been observed that at all {\ppt} regions, HIJING and Sibyll model do not show strong rapidity ($y$) dependence at all. Therefore, the ratio is about unity which means that in HIJING and Sibyll, the same number of {\alam} are produced as that of {\lam}. However, the QGSJET model describes the distribution of {\alam}/{\lam} ratios reasonably well. At $0.65 < p_T < 1.0$ GeV/$c$ region, below $y < 3$, QGSJET slightly overpredict the experimental results but at high $y$ region it is in good agreement with the experimental observations. Overall, QGSJET is in good agreement with data at all the {\ppt} regions.

Figure~\ref{fig3} (right column) depict the model prediction of baryon to meson ({\lam}/{\ks}) ratios as a function of rapidity ($y$) compared with the experimental results. At all {\ppt} regions, it can bee seen that, again the predictions of HIJING and Sibyll shows no strong rapidity ($y$) dependence. However, HIJING predicts the experimental results reasonably well as compared to the Sibyll and QGSJET. Sibyll completely underpredict the experimental as well as HIJING and QGSJET results at $0.25 < p_T < 0.65$ GeV/$c$ and $0.65 < p_T < 1.0$ GeV/$c$ regions. QGSJET slightly underpredict the experimental data at $0.25 < p_T < 0.65$ GeV/$c$ where gives a reasonable description at $0.65 < p_T < 1.0$ GeV/$c$ region. At $1.0 < p_T < 2.5$ GeV/$c$, all of the model prediction are roughly similar and are consistent with experimental data.

\begin{figure}[ht!]
\centering
\begin{subfigure}[b]{0.49\textwidth}
\centering
\includegraphics[width=\textwidth]{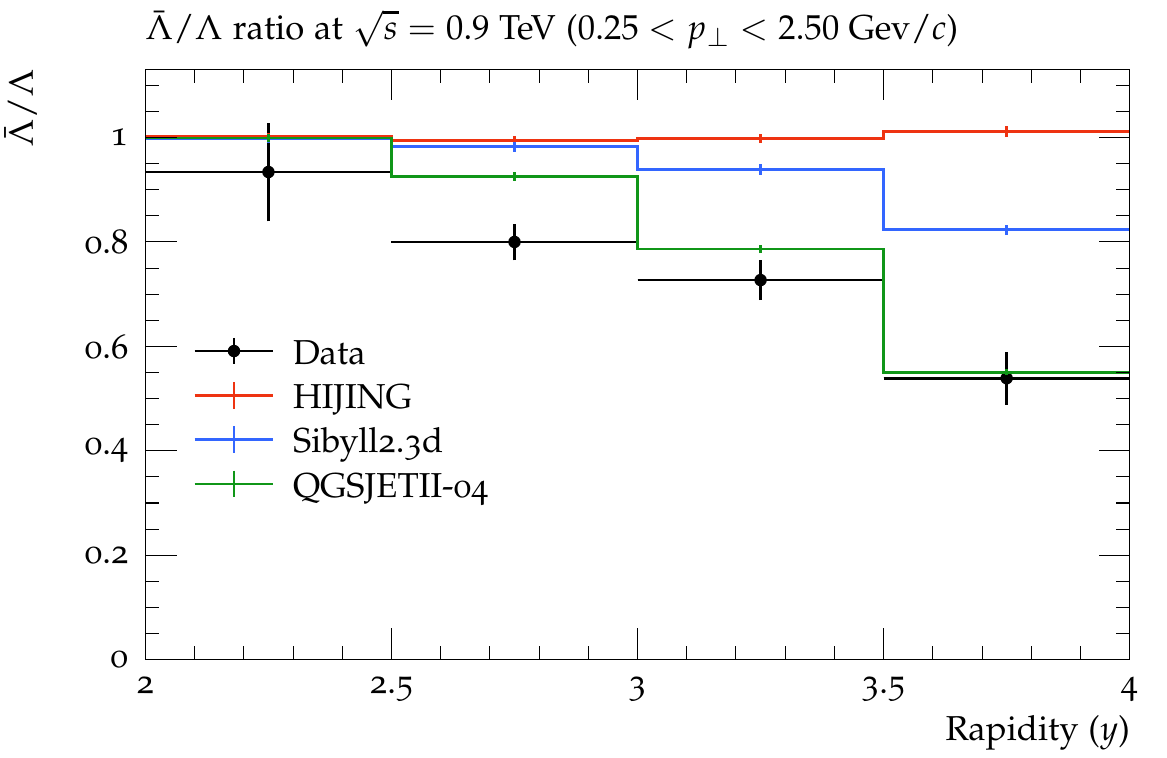}
\caption{{\alam}/{\lam} Ratio as a function of rapidity ($y$)}
\label{fig4a}
\end{subfigure}
\hfill
\begin{subfigure}[b]{0.49\textwidth}
\centering
\includegraphics[width=\textwidth]{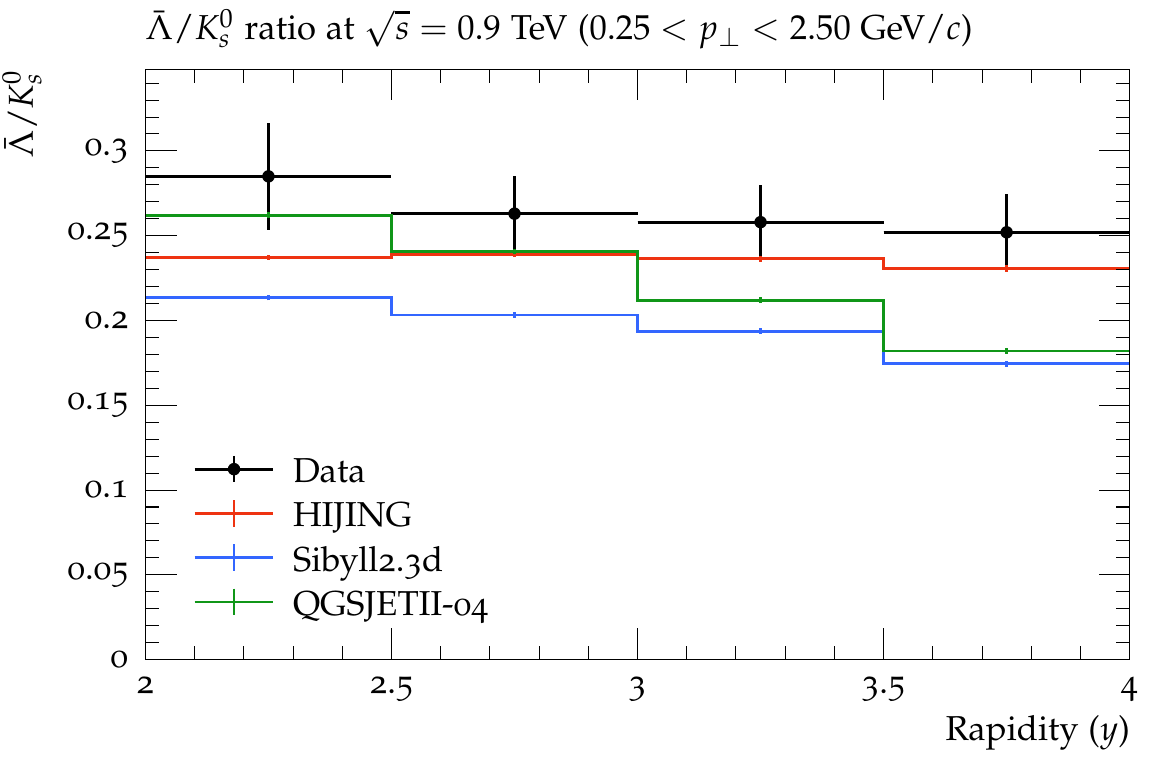}
\caption{{\alam}/{\ks} Ratio as a function of rapidity ($y$)}
\label{fig4b}
\end{subfigure}
\hfill
\begin{subfigure}[b]{0.49\textwidth}
\centering
\includegraphics[width=\textwidth]{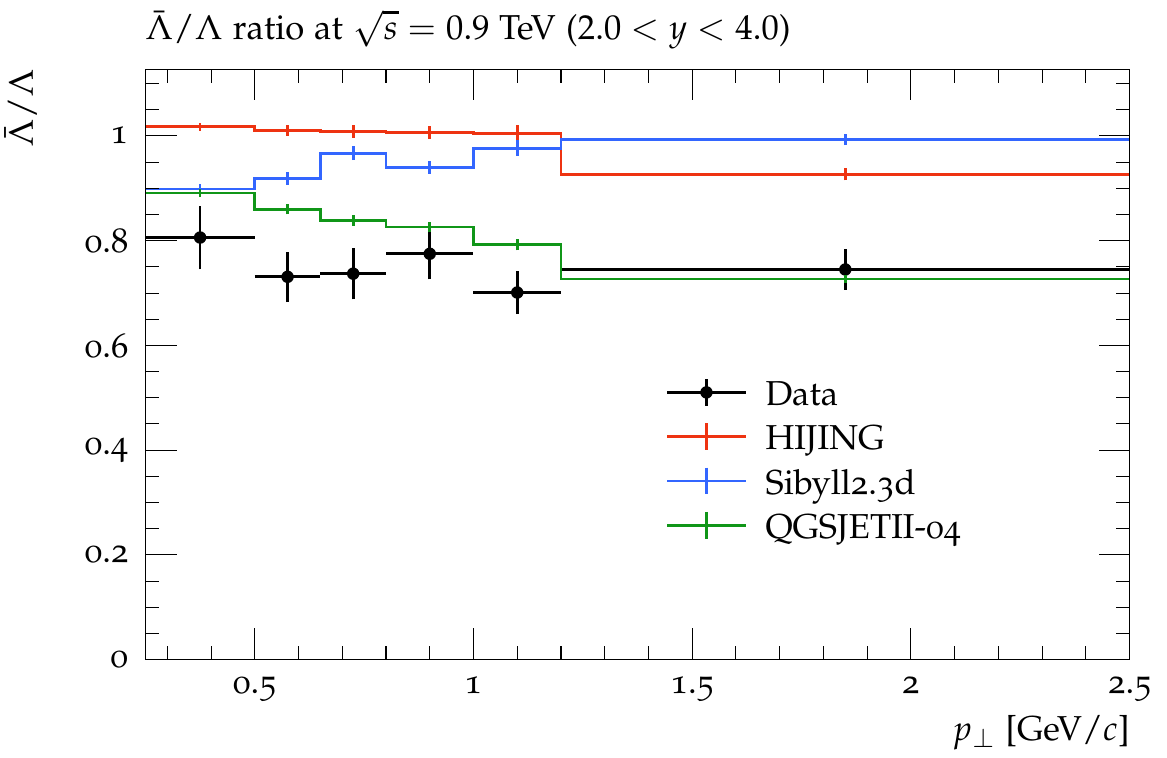}
\caption{{\alam}/{\lam} Ratio as a function of {\ppt}}
\label{fig4c}
\end{subfigure}
\hfill
\begin{subfigure}[b]{0.49\textwidth}
\centering
\includegraphics[width=\textwidth]{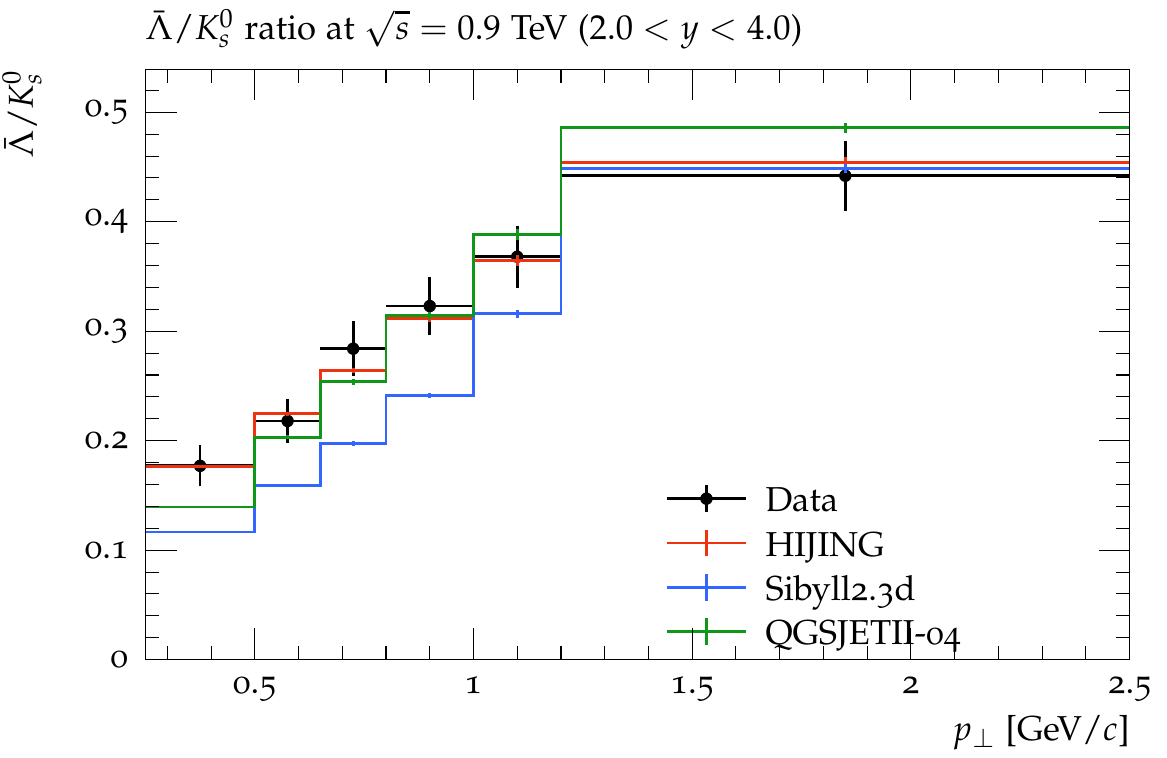}
\caption{{\alam}/{\ks} Ratio as a function of {\ppt}}
\label{fig4d}
\end{subfigure}
\hfill
\begin{subfigure}[b]{0.49\textwidth}
\centering
\includegraphics[width=\textwidth]{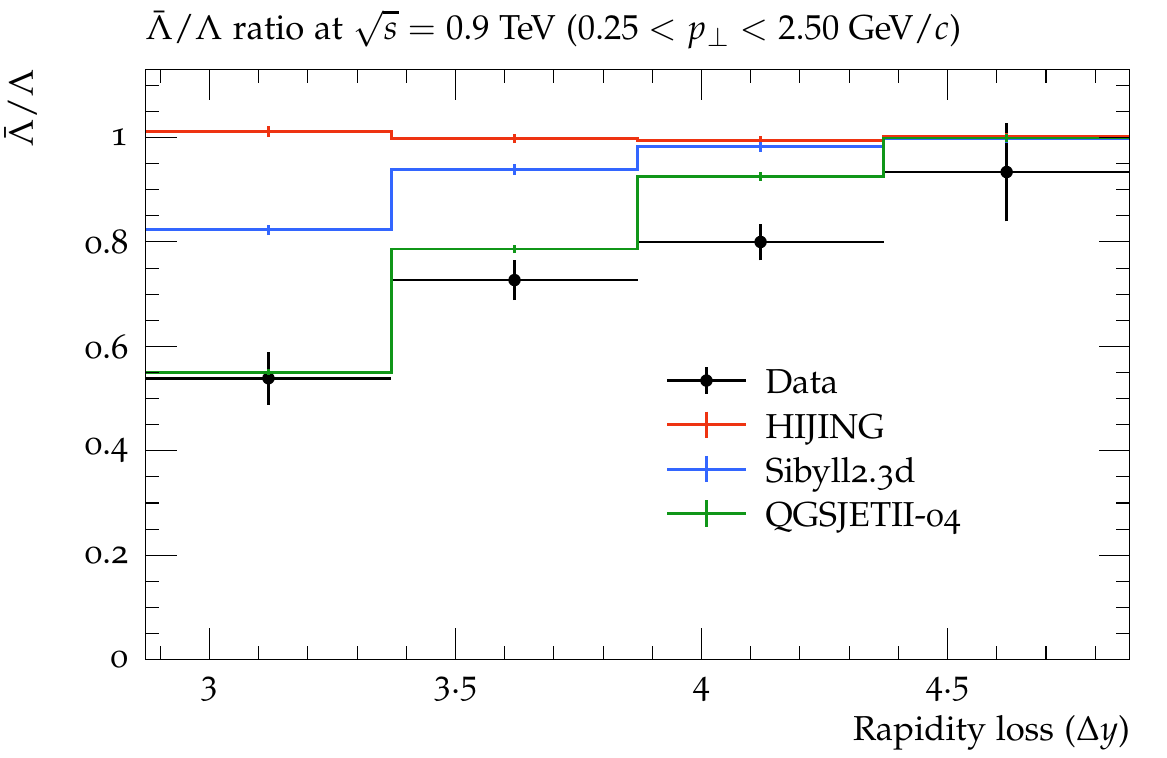}
\caption{{\alam}/{\lam} Ratio as a function of rapidity loss ($\Delta y$)}
\label{fig4e}
\end{subfigure}
\hfill
\begin{subfigure}[b]{0.49\textwidth}
\centering
\includegraphics[width=\textwidth]{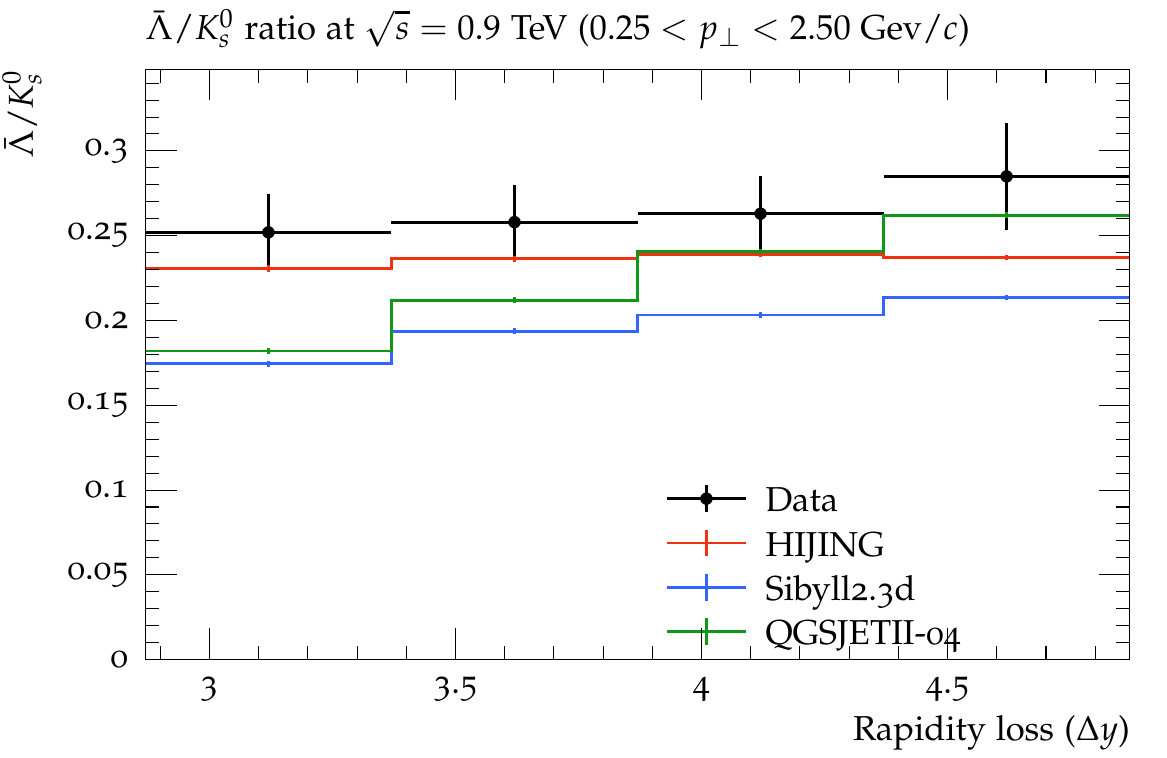}
\caption{{\alam}/{\ks} Ratio as a function of rapidity loss ($\Delta y$)}
\label{fig4f}
\end{subfigure}
\caption{Different ratios as a function of {\ppt}, rapidity ($y$) and rapidity loss ($\Delta y$) in $pp$ collisions at \sqrts~ = 0.9 TeV from LHCb experiment in comparison to the model predictions. Black solid markers are the data points and lines of different colors shows different model predictions. }
\label{fig4}
\end{figure}

We have also combined all the {\ppt} regions from $0.25 < p_T < 2.5$ GeV/$c$ and studied the models predictions of various ratios as a function of {\ppt}, rapidity ($y$) and rapidity loss ($\Delta y$) in comparison with experimental results. Figure~\ref{fig4a} shows the {\alam}/{\lam} ratio as a function of rapidity ($y$) at all {\ppt} range. Only QGSJET model is consistent with experimental data while HIJING and Sibyll does not show strong rapidity dependence. {\alam}/{\ks} ratio as a function of rapidity ($y$) is shown in fig.~\ref{fig4b}. HIJING results are slightly close to the data and does not show rapidity dependence. On the other hand, Sibyll and QGSJET underpredict the experimental data and HIJING data significantly. Figure~\ref{fig4c} depict the {\alam}/{\lam} ratio as a function of {\ppt} in the rapidity range $2.0 < y < 4.0$. Only QGSJET agrees with the data at {\ppt} $> 0.8$ GeV/$c$, while HIJING and Sibyll significantly overshoot the data.

HIJING does not show {\ppt} dependence at all, while Sibyll and QGSJET do. {\alam}/{\ks} ratio as a function of {\ppt} can be seen in fig.~\ref{fig4d}. QGSJET and HIJING reasonably describe the data as well as the ratio distribution at all {\ppt} regions, while Sibyll slightly underpredicts the data at almost all the {\ppt} regions. Figures~\ref{fig4e} and ~\ref{fig4f} respectively shows the {\alam}/{\lam} and {\alam}/{\ks} ratios as a function of rapidity loss ($\Delta y$). In the case of {\alam}/{\lam} in fig~\ref{fig4e}, only QGSJET describes the data and distribution while HIJING and Sibyll overshoot the data. HIJING also does not shows the $\Delta y$ dependence at all. {\alam}/{\ks} ratios from fig.~\ref{fig4f} almost all of the models underpredict the data significantly except HIJING, whose predictions are close to the experimental observations.

\begin{figure}[ht!]
\centering
\includegraphics[width=0.49\textwidth]{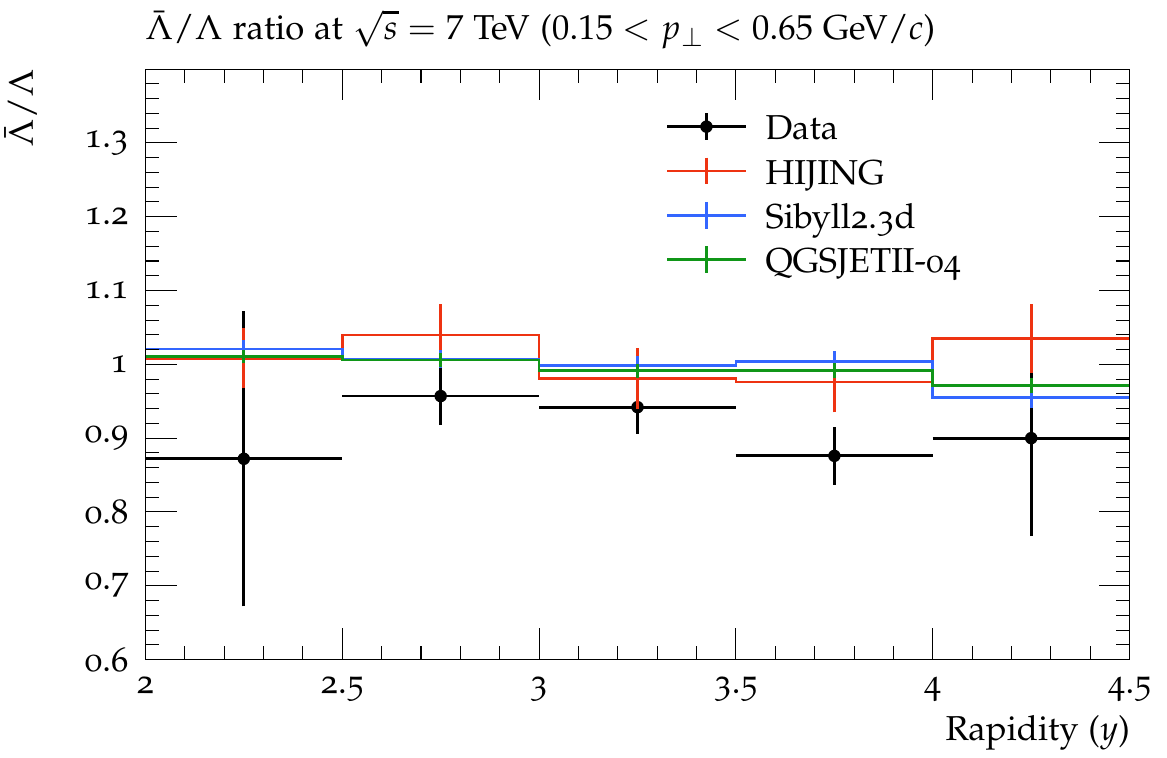}
\includegraphics[width=0.49\textwidth]{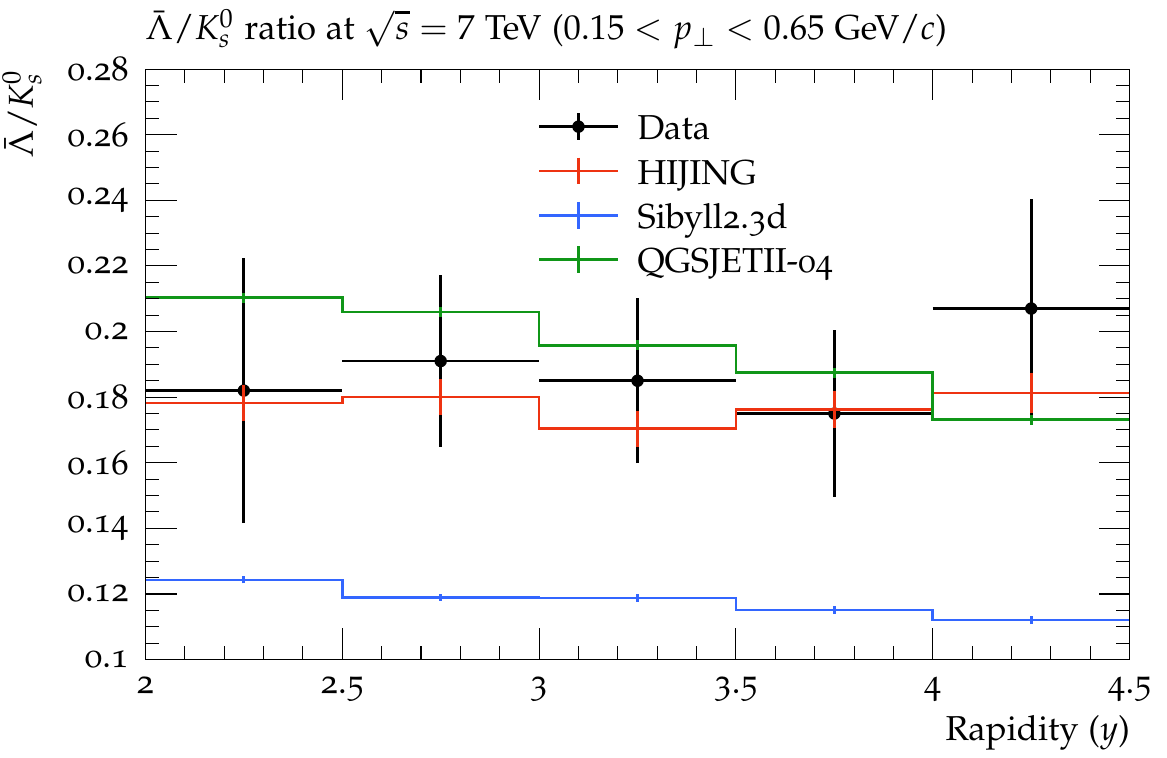}
\includegraphics[width=0.49\textwidth]{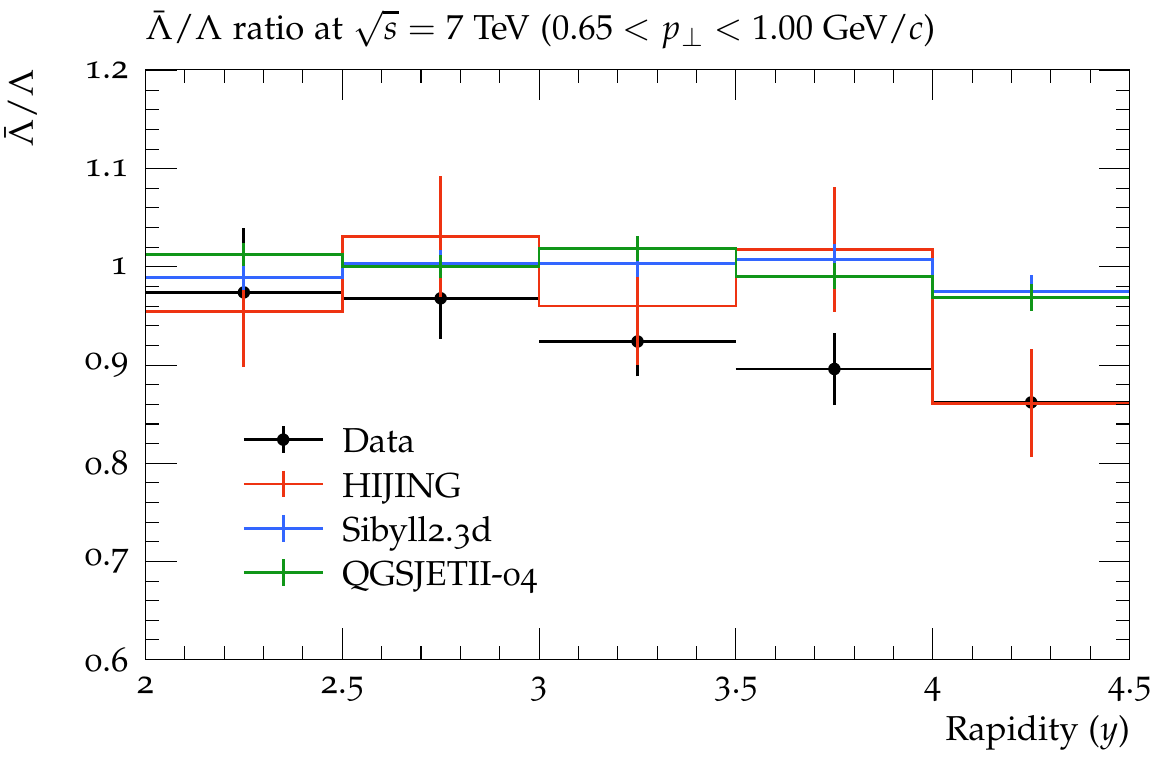}
\includegraphics[width=0.49\textwidth]{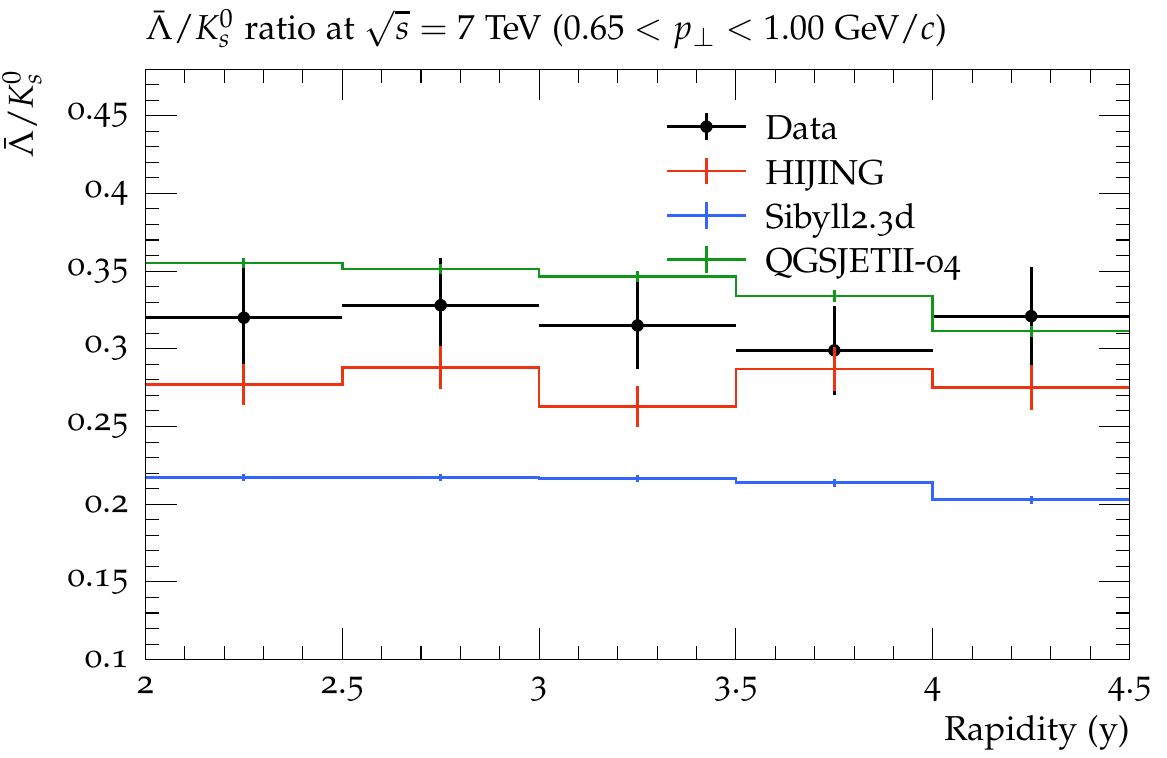}
\includegraphics[width=0.49\textwidth]{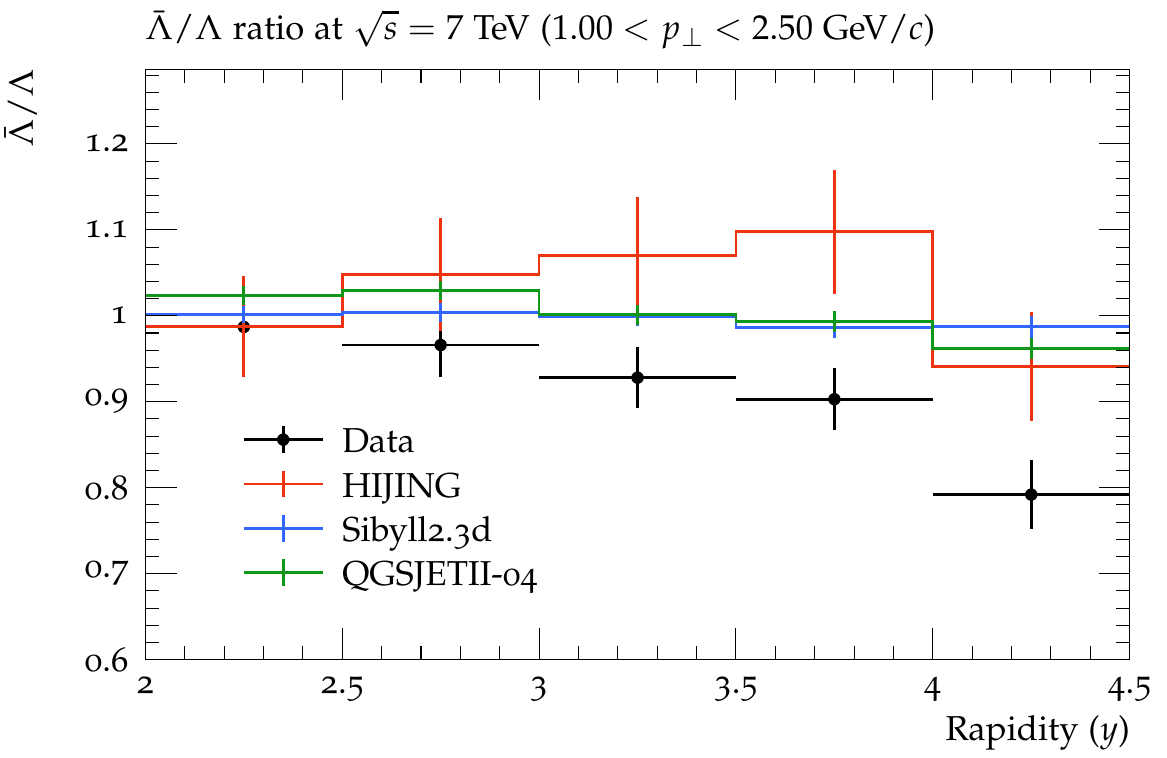}
\includegraphics[width=0.49\textwidth]{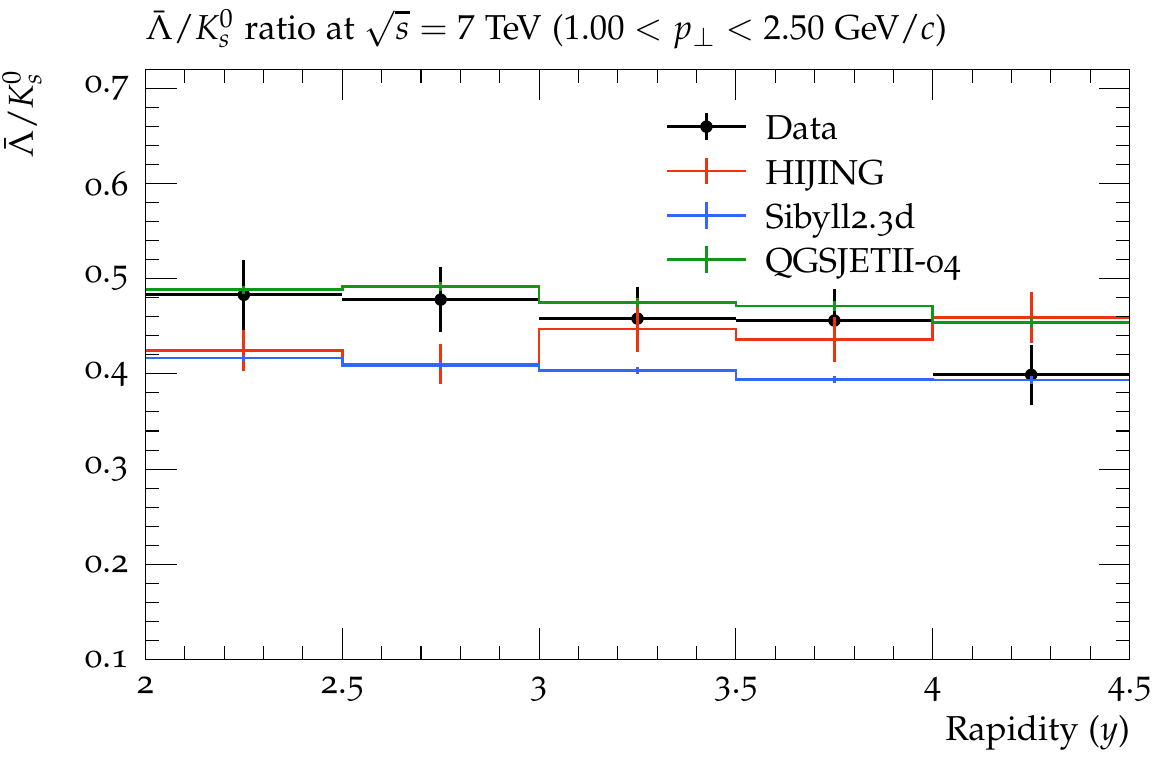}
\caption{(Left column) anti-baryon to baryon ({\alam}/{\lam}) ratios as a function of rapidity ($y$) (Right Column) anti-baryon to meson ({\alam}/{\ks}) ratios in $pp$ collisions \sqrts~ = 7 TeV at different {\ppt} regions. Black solid markers are the data points and lines of different colors shows different model predictions.}
\label{fig5}
\end{figure}

For the comparison study of various particle ratios in $pp$ collisions at \sqrts~ = 7 TeV, the {\ppt} is divided into various regions; $0.15 < p_T < 0.65$ GeV/$c$, $0.65 < p_T < 1.0$ GeV/$c$ and $1.0 < p_T < 2.5$ GeV/$c$ in the rapidity $2 < y < 4$ region. The anti-baryon to baryon ({\alam}/{\lam}) and baryon to meson ({\alam}/{\ks}) ratios are then compared with different model predictions at the given {\ppt} and $y$ regions.

Figure~\ref{fig5} (left column) shows the model prediction of anti-baryon to baryon ({\alam}/{\lam}) ratios as a function of rapidity ($y$) in comparison to the data from LHCb experiment in $pp$ collisions at \sqrts~ = 7 TeV~\cite{22}. A slight $y$ dependence can be observed from the experimental measurements. At $0.15 < p_T < 0.65$ GeV/$c$ region, the ratio predictions of the models is close to unity while experimental data lies under the model predictions. At $0.65 < p_T < 1.0$ GeV/$c$ region, Sibyll and QGSJET predictions are close to unity, while a small deviation is observed in HIJING predictions. The HIJING slightly describes the shape of the ratio distribution. The HIJING model at $1.0 < p_T < 2.5$ GeV/$c$ region behave differently and overshoot the data and hence does not describe the distribution shape correctly. On the other hand, Sibyll and QGSJET ratio is close to unity. Overall, none of the models describe the experimental data and distribution shape accurately at all {\ppt} regions.

Figure~\ref{fig5} (right column) depict the model prediction of baryon to meson ({\alam}/{\ks}) ratios as a function of rapidity ($y$) compared with the experimental results. At $0.15 < p_T < 0.65$ GeV/$c$ region, HIJING and QGSJET describe the experimental data within uncertainties; however, Sibyll completely fails to predict the experimental results and hence underpredict the data. At $0.65 < p_T < 1.0$ GeV/$c$ region, QGSJET predictions match with data within uncertainties while HIJING slightly undershoots the data. Sibyll completely fails to predict the experimental results and undershoot the experimental observations. At $1.0 < p_T < 2.5$ GeV/$c$ region, Sibyll predictions are slightly closer to the HIJING but still lower than HIJING and QGSJET predictions and hence experimental data. Only the QGSJET model completely describes the experimental observations.

\begin{figure}[ht!]
\centering
\begin{subfigure}[b]{0.49\textwidth}
\centering
\includegraphics[width=\textwidth]{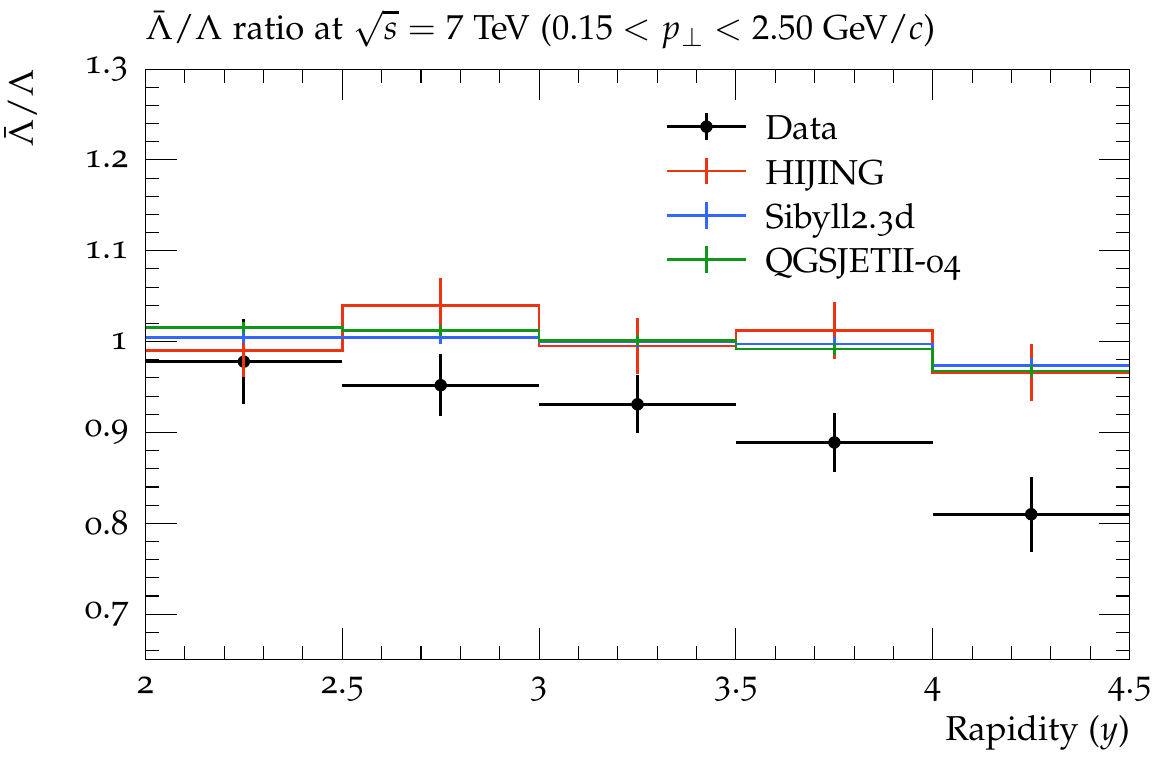}
\caption{{\alam}/{\lam} Ratio as a function of rapidity ($y$)}
\label{fig6a}
\end{subfigure}
\hfill
\begin{subfigure}[b]{0.49\textwidth}
\centering
\includegraphics[width=\textwidth]{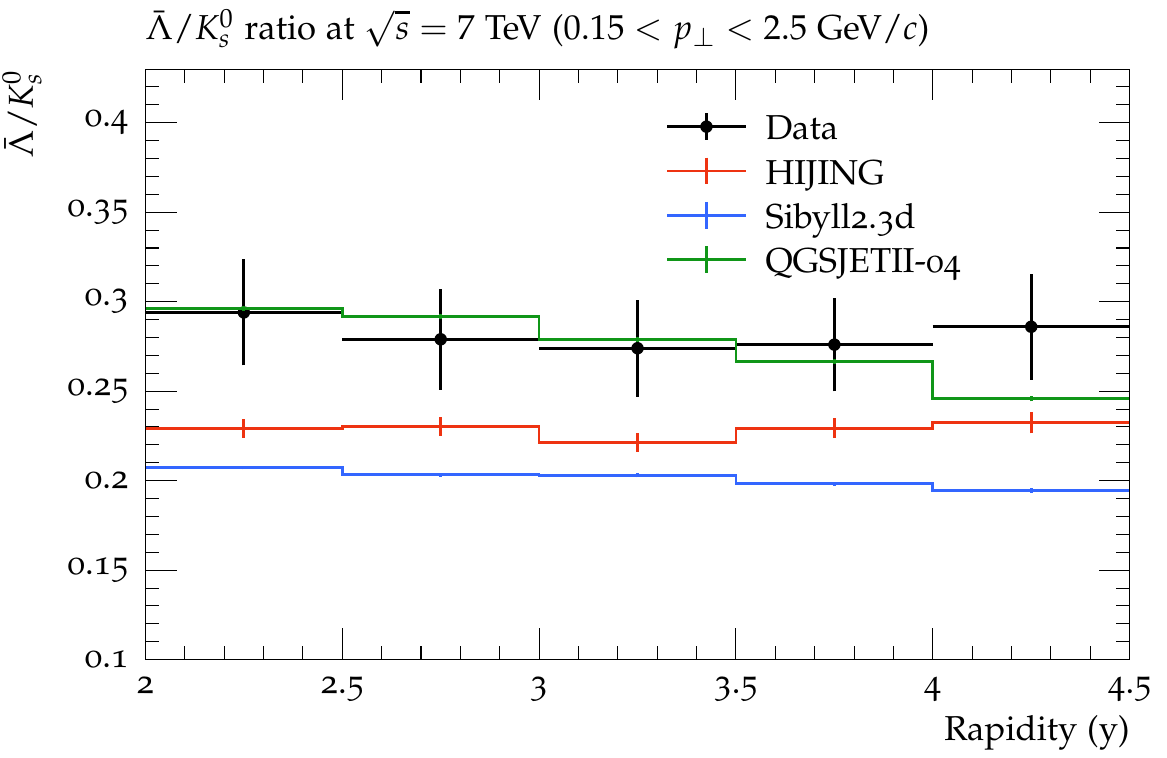}
\caption{{\alam}/{\ks} Ratio as a function of rapidity ($y$)}
\label{fig6b}
\end{subfigure}
\hfill
\begin{subfigure}[b]{0.49\textwidth}
\centering
\includegraphics[width=\textwidth]{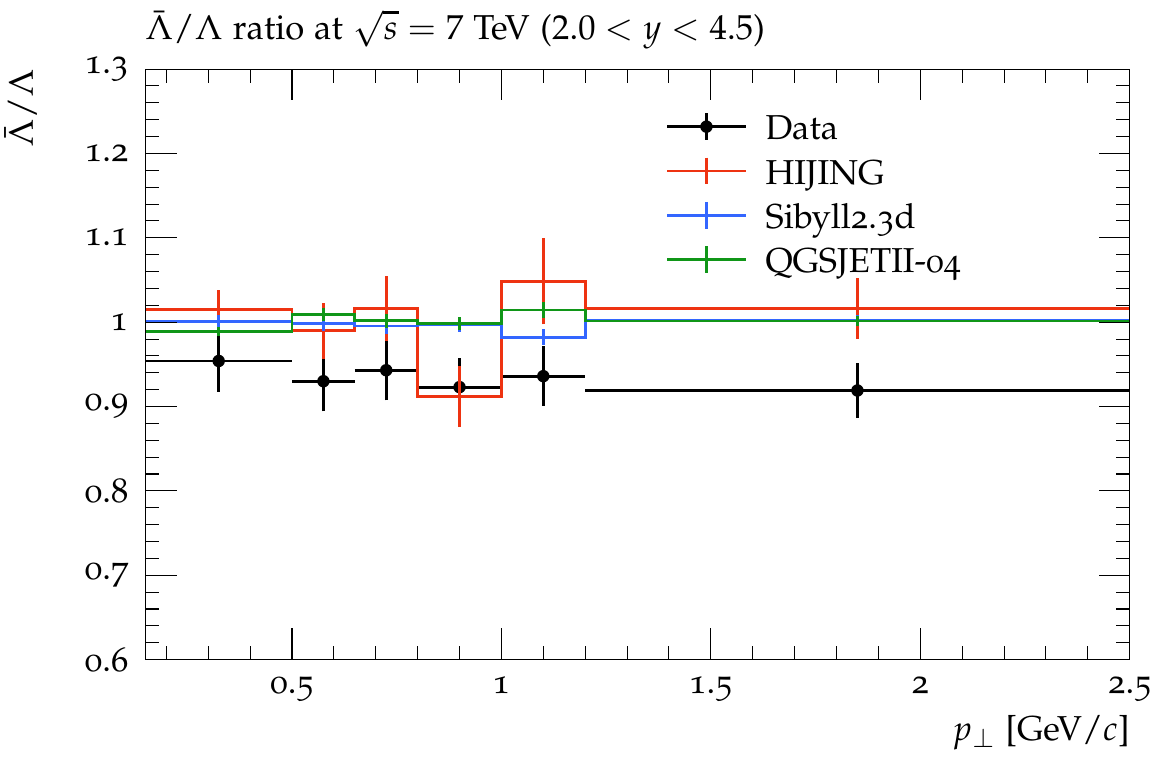}
\caption{{\alam}/{\lam} Ratio as a function of {\ppt}}
\label{fig6c}
\end{subfigure}
\hfill
\begin{subfigure}[b]{0.49\textwidth}
\centering
\includegraphics[width=\textwidth]{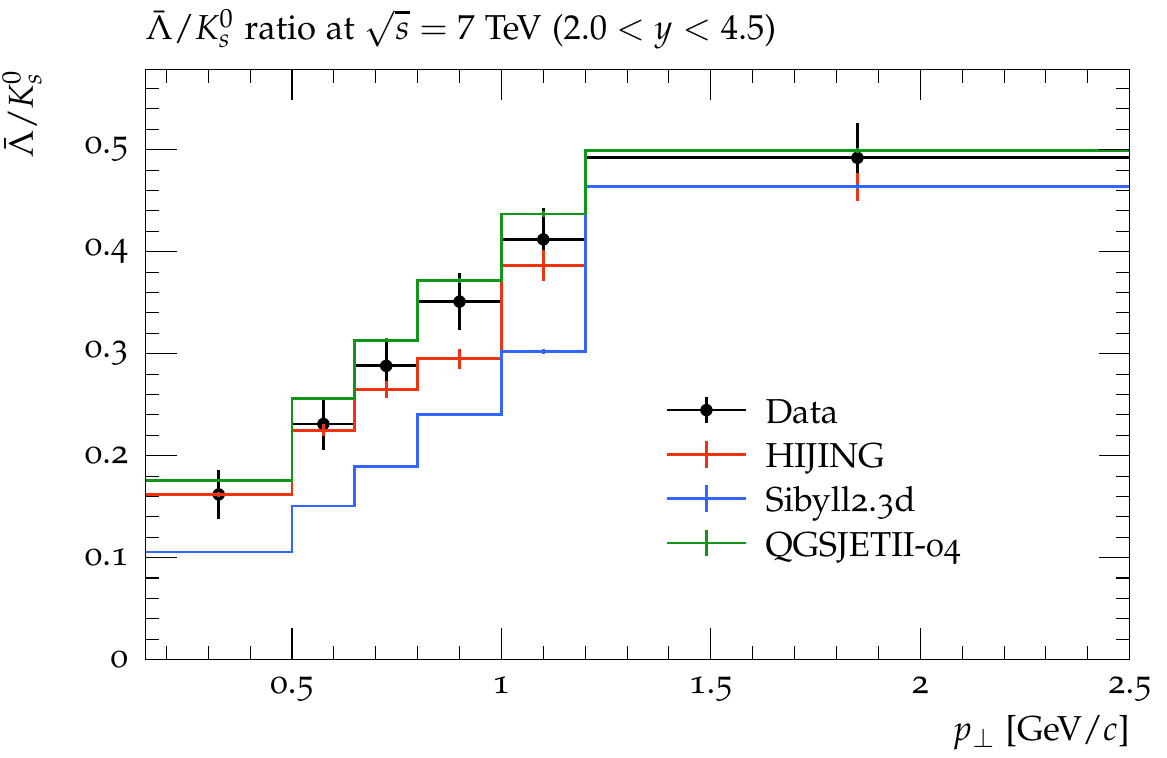}
\caption{{\alam}/{\ks} Ratio as a function of {\ppt}}
\label{fig6d}
\end{subfigure}
\hfill
\begin{subfigure}[b]{0.49\textwidth}
\centering
\includegraphics[width=\textwidth]{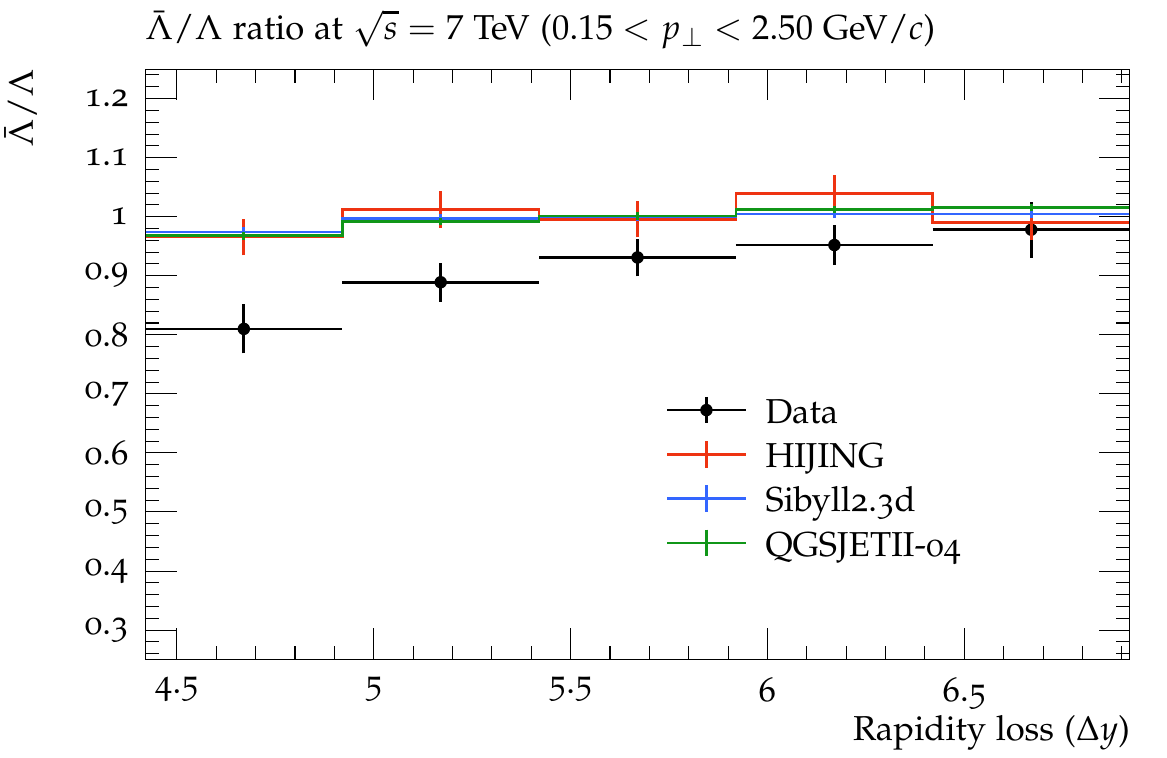}
\caption{{\alam}/{\lam} Ratio as a function of rapidity loss ($\Delta y$)}
\label{fig6e}
\end{subfigure}
\hfill
\begin{subfigure}[b]{0.49\textwidth}
\centering
\includegraphics[width=\textwidth]{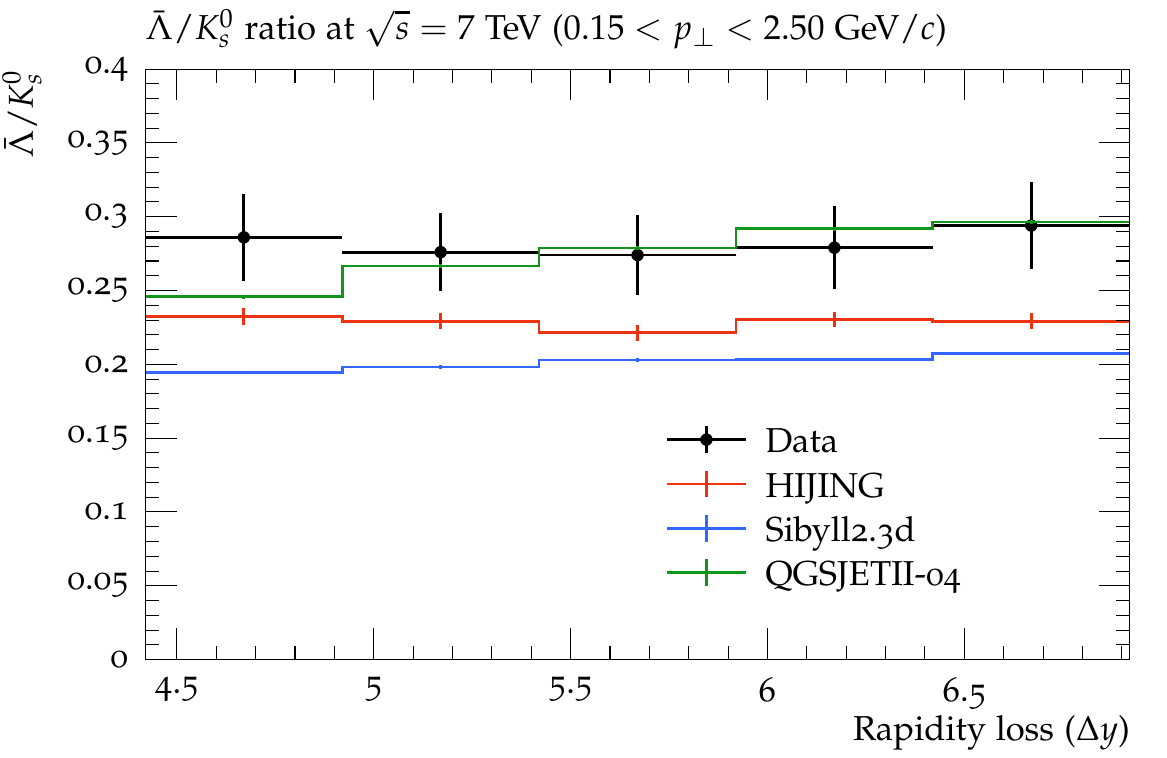}
\caption{{\alam}/{\ks} Ratio as a function of rapidity loss ($\Delta y$)}
\label{fig6f}
\end{subfigure}
\caption{Different ratios as a function of {\ppt}, rapidity ($y$) and rapidity loss ($\Delta y$) in $pp$ collisions at \sqrts~ = 7 TeV from LHCb experiment in comparison to the model predictions. Black solid markers are the data points and lines of different colors shows different model predictions. }
\label{fig6}
\end{figure}

Figure~\ref{fig6a} shows the {\alam}/{\lam} ratio as a function of rapidity ($y$) at all {\ppt} range from $0.15 < p_T < 2.50$ GeV/$c$ region. All the model predictions are close to unity and roughly about the same, while data lies below the model predictions and shows the decreasing trend with rapidity $y$. All the models do not show rapidity dependence at all. The {\alam}/{\ks} ratio as a function of rapidity ($y$) is shown in fig.~\ref{fig6b}. Only the prediction of QGSJET is in good agreement with the data, while HIJING and Sibyll's predictions are undershooting the data and QGSJET significantly. Figure~\ref{fig6c} depict the {\alam}/{\lam} ratio as a function of {\ppt} in the rapidity range $2.0 < y < 4.0$. None of the models describe the experimental data. The model predictions are close to unity while data lies below around 0.95. Both data and all the models do not show {\ppt} dependence. {\alam}/{\ks} ratio as a function of {\ppt} can be seen in fig.~\ref{fig6d}. QGSJET and HIJING reasonably describe the data within uncertainties as well as the ratio distribution at all {\ppt} regions. At the same time, Sibyll slightly underpredicts the data at almost all the {\ppt} regions. Figures~\ref{fig6e} and ~\ref{fig6f} respectively shows the {\alam}/{\lam} and {\alam}/{\ks} ratios as a function of rapidity loss ($\Delta y$). In the case of {\alam}/{\lam} in fig~\ref{fig4e}, all of the model predictions are close to unity, and data lies significantly lower than the model data. None of the models explains this distribution well. {\alam}/{\ks} ratios from fig.~\ref{fig6a} almost all of the models underpredict the data significantly except QGSJET, whose predictions are consistent with the experimental observations.

\begin{figure}[ht!]
\centering
\begin{subfigure}[b]{0.49\textwidth}
\centering
\includegraphics[width=\textwidth]{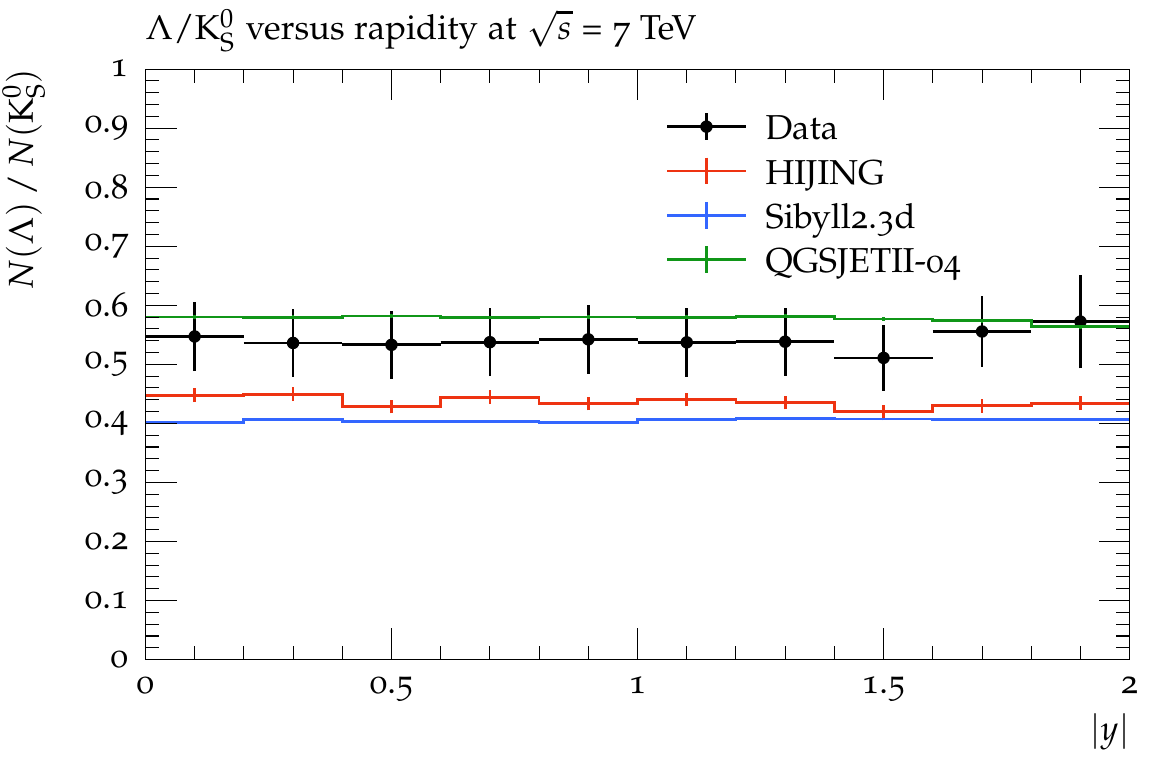}
\caption{{\lam}/{\ks} Ratio as a function of rapidity ($|y|$)}
\label{fig7a}
\end{subfigure}
\hfill
\begin{subfigure}[b]{0.49\textwidth}
\centering
\includegraphics[width=\textwidth]{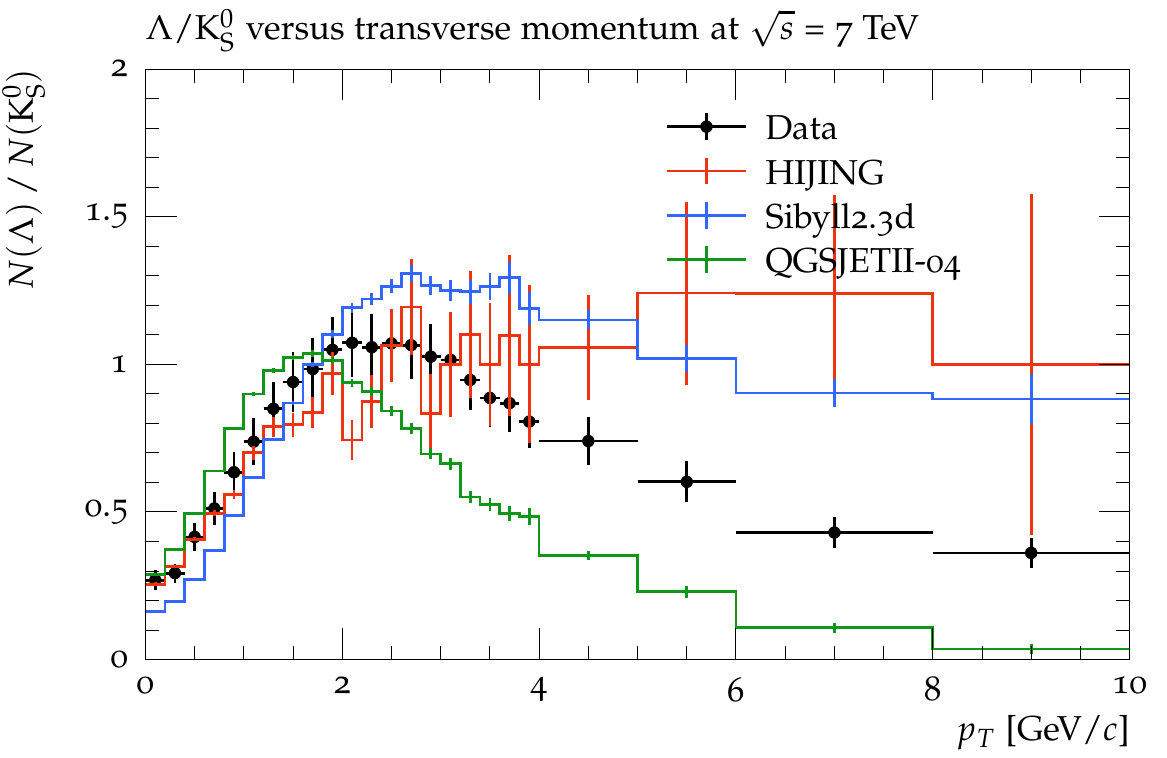}
\caption{{\lam}/{\ks} Ratio as a function of {\ppt}}
\label{fig7b}
\end{subfigure}
\hfill
\begin{subfigure}[b]{0.49\textwidth}
\centering
\includegraphics[width=\textwidth]{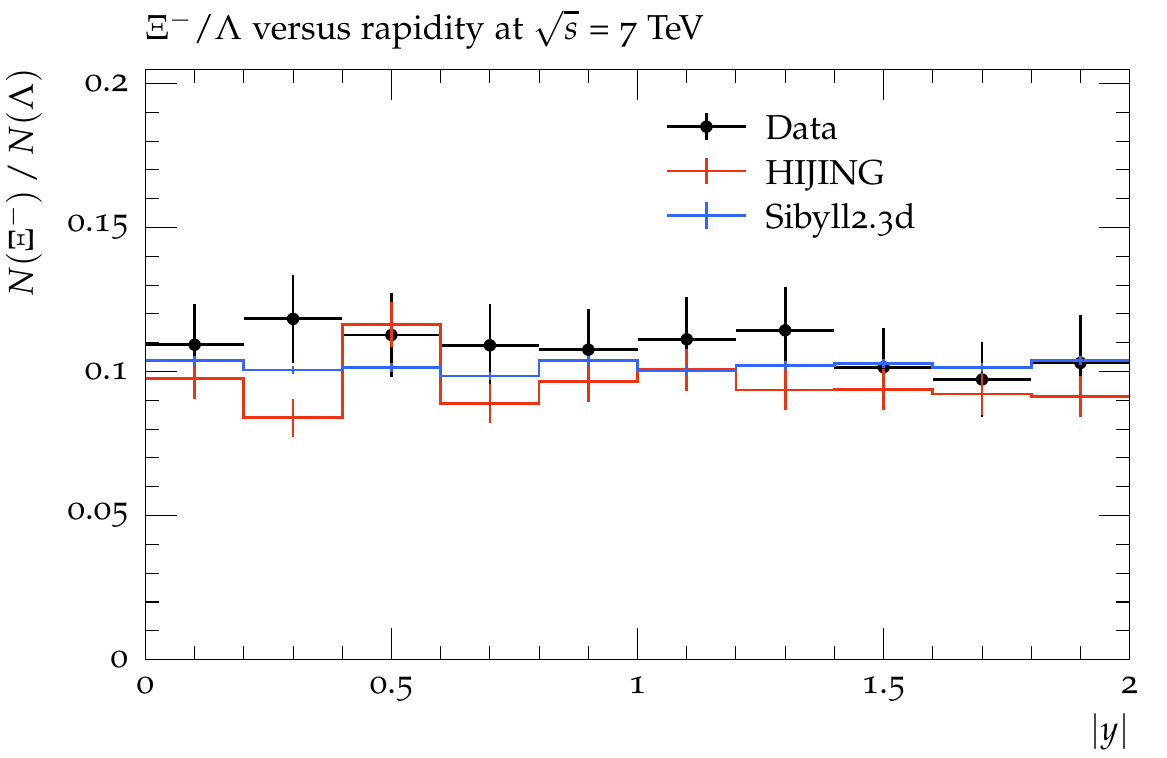}
\caption{{\xim}/{\lam} Ratio as a function of rapidity ($|y|$)}
\label{fig7c}
\end{subfigure}
\hfill
\begin{subfigure}[b]{0.49\textwidth}
\centering
\includegraphics[width=\textwidth]{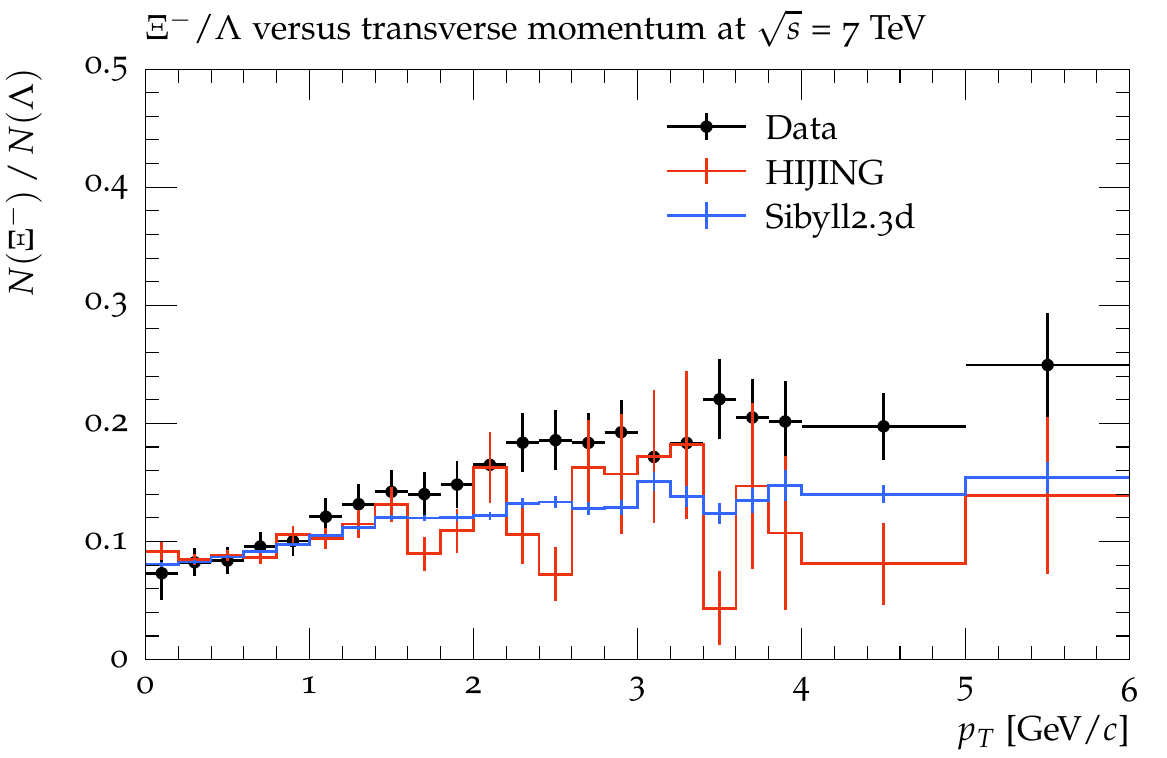}
\caption{{\xim}/{\lam} Ratio as a function of {\ppt}}
\label{fig7d}
\end{subfigure}
\caption{Different ratios as a function of {\ppt} and rapidity ($y$) in $pp$ collisions at \sqrts~ = 7 TeV from LHCb experiment in comparison to the model predictions. Black solid markers are the data points and lines of different colors shows different model predictions. }
\label{fig7}
\end{figure}

Figure~\ref{fig7a} shows the {\lam}/{\ks} ratio as a function of {\ppt} from $pp$ collisions at \sqrts~ = 7 TeV in comparison with the experimental results. The data shows no rapidity $y$ dependence and almost follows the straight line. It can also be seen that all of the models also do not show the rapidity $y$ dependence at all, but QGSJET is the only model which predicts the same results as that of data within uncertainties. Sibyll and HIJING predictions are very close but underpredict the experimental data and QGSJET roughly by 15\% and 10\%, respectively. Figure~\ref{fig7b} presents the {\lam}/{\ks} ratio as a function of {\ppt}. It is observed that there is sharper increases in {\lam} spectrum with the increase in low {\ppt} ({\ppt} $< 1.5$ GeV/$c$) and {\lam} production is relatively larger at the intermediate {\ppt} region ($1.5 \le$ {\ppt} $\le 4.0$ GeV/$c$). The same shape difference between {\lam} and {\ks} can be seen as reported in baryon to meson ratios~\cite{25}. HIJING is observed to predict the almost similar shape and experimental results from low to intermediate {\ppt} regions, i,e, up to {\ppt} $\le 2.5$ GeV/$c$. Sibyll slightly underpredict the data up to {\ppt} $< 1.5$ GeV/$c$ and overshoot at higher {\ppt} regions. QGSJET, on the other hand, slightly overpredicts the experimental and both model data up to {\ppt} $< 1.5$ GeV/$c$ and undershoot at above {\ppt} regions. This difference in the shape of experimental results and model predictions may be related to the fact that there is a significant contribution from the fragmentation processes to meson production compared to the baryon production, which is based on mass and energy arguments. 
Furthermore, the measurements of {\xim}/{\lam} ratios have also been performed with these models. Figure~\ref{fig7c} shows the {\xim}/{\lam} ratio as a function of rapidity $y$. The QGSJET model does not include the definition of $\Xi$ particle; therefore, no comparison can be made. However, HIJING and Sibyll predict similar results to the experimental data within uncertainties. Figure~\ref{fig7d} presents {\xim}/{\lam} ratios as a function of {\ppt}. Both models Sibyll and HIJING describe the data up to {\ppt} $< 1.5$ GeV/$c$, while underpredict the data at higher {\ppt} regions. Sibyll shows balanced distribution at higher {\ppt} regions while fluctuation is observed in HIJING predictions which may be due to statistics. 
For a quantitative description of the particles' ratio, we have calculated the values of $\chi^2/n$ for all models for the energies above. The Values are tabulated in Table 1, where the first column shows the power, followed by the figures and particles ratios in columns 2 and 3, respectively. The following three columns show the values of the $\chi^2$/$n$ for the HIJING, Sibyll2.3d, and QGSJETII-04 models, respectively. The last column shows the numerical values of $n$, where $n$ indicates the number of measured points on the x-axis in each case. 
Taking into account the statistical error bars in the models' prediction and the values of the $\chi^2$/$n$ for Fig. 1, it is clear that the models are in good agreement with the experimental data, while large values of $\chi^2$/$n$ for Fig. 2 show that 
models failed to reproduce the measurements. At 0.9 and 7 TeV, the QGSJET model has the lowest values of $\chi^2$/$n$ consistently compared to the other two models. For different particle ratios ({\alam}/{\ks}, {\xim}/{\lam}) the HIJING model performed better even than the QGSJET model but failed completely to prediction the same particles ratio ({\alam}/{\lam}) at 0.9 and 7 TeV. The values of $\chi^2$/$n$ for the different particles' ratios in the Sibyll model are lower than the HIJING but comparatively higher than the QGSJET model. Still, they have very high values for the same particles' ratios.

\begin{table*}[hbt!]
{\scriptsize Table 1. The values of the $\chi^2/n$ for the HIJING, Sibyll2.3d and QGSJETII-04 models at $\sqrt{s}$ = 0.2, 0.9 and 7 TeV.
\begin{center}
\begin{tabular} {cccccccccccc}\\ \hline\hline
&$Energy$ &$Figure$ & ratio & $\chi^2/n$ & $\chi^2/n$ & $\chi^2/n$\\
&&&& $HJIJING$ & $Sibyll2.3d$ & $QJSJETII-04$ & $n$\\
\hline
& & 1(a) & {\alam}/{\lam} vs {\ppt} & 2.29 & 1.78 & 1.99 & 21\\
& 200 GeV & 1(b) & {\axi}/{\xim} vs {\ppt} & 0.99 & 0.71 & - & 11\\
& & 2 & $\bar B/ B$ vs strangness & 18.75 & 63.23 & 75.76 & 4\\
\hline
&& 3(a) & {\alam}/{\lam} vs $y$ & 18.86 & 8.19 & 5.62 & 4\\
&& 3(b) & {\alam}/{\ks} vs $y$ & 0.52 & 5.00 & 1.85 & 4\\
&900& 3(c) & {\alam}/{\lam} vs $y$ & 22.10 & 12.72 & 4.40 & 4\\
&GeV& 3(d) & {\alam}/{\ks} vs $y$ & 0.48 & 9.62 & 1.56 & 4\\
&& 3(e) & {\alam}/{\lam} vs $y$ & 23.43 & 30.52 & 2.41 & 4\\
&& 3(f) & {\alam}/{\ks} vs $y$ & 1.70 & 0.67 & 1.45 & 4\\
\hline
&& 4(a) & {\alam}/{\lam} vs $y$ & 39.88 & 20.92 & 3.70 & 4\\
&& 4(b) & {\alam}/{\ks} vs $y$ & 1.32 & 8.12 & 3.83 & 6\\
&900& 4(c) & {\alam}/{\lam} vs {\ppt} & 26.88 & 20.81 & 3.14 & 6\\
&GeV& 4(d)& {\alam}/{\ks} vs {\ppt} & 0.18 & 7.22 & 1.41 & 6\\
&& 4(e) & {\alam}/{\lam} vs $\Delta y$ & 39.88 & 20.92 & 3.70 & 4\\
&& 4(f) & {\alam}/{\ks} vs $\Delta y$ & 1.32 & 8.12 & 3.83 & 4\\
\hline

&& 5(a) & {\alam}/{\lam} vs $y$ & 1.40 & 2.74 & 2.41 & 5\\
&& 5(b) & {\alam}/{\ks} vs $y$ & 0.21 & 6.03 & 0.45 & 5\\
&7000& 5(c) & {\alam}/{\lam} vs $y$ & 0.76 & 3.43 & 3.39 & 5\\
&GeV& 5(d) & {\alam}/{\ks} vs $y$ & 1.52 & 11.58 & 0.90 & 5\\
&& 5(e) & {\alam}/{\lam} vs $y$ & 2.86 & 6.17 & 5.81 & 5\\
&& 5(f) & {\alam}/{\ks} vs $y$ & 1.44 & 2.71 & 0.71 & 5\\
\hline
&& 6(a) & {\alam}/{\lam} vs $y$ & 4.44 & 6.48 & 6.36 & 5\\
&& 6(b) & {\alam}/{\ks} vs $y$ & 3.47 & 8.20 & 0.44 & 5\\
&7000& 6(c) & {\alam}/{\lam} vs {\ppt} & 2.13 & 3.19 & 3.89 & 6\\
&GeV& 6(d) & {\alam}/{\ks} vs {\ppt} & 0.90 & 9.60 & 0.57 & 6\\
&& 6(e) & {\alam}/{\lam} vs $\Delta y$ & 4.44 & 6.48 & 6.36 & 5\\
&& 6(f) & {\alam}/{\ks} vs $\Delta y$ & 3.47 & 8.20 & 0.44 & 5\\
\hline

&& 7(a) & {\lam}/{\ks} vs $|y|$ & 2.97 & 5.17 & 0.52 & 10\\
&& 7(b) & {\lam}/{\ks} vs {\ppt} & 1.29 & 9.67 & 9.66 & 24\\
&7000& 7(c) & {\xim}/{\lam} vs $|y|$ & 0.97 & 0.47 & - & 10\\
&GeV& 7(d) & {\xim}/{\lam} vs {\ppt} & 2.36 & 2.47 & - & 22\\

\hline
\end{tabular}%
\end{center}}
\end{table*}

\section{Conclusion}\label{sec4}

We have performed systematic study of different particle ratios {\alam}/{\lam}, {\alam}/{\ks} and {\xim}/{\lam} as a function of rapidity $y$ and {\ppt} from $pp$ collisions at \sqrts~= 0.2, 0.9, and 7 TeV using Sibyll, HIJING and QGSJET and detailed comparison has been made with available experimental data. The anti-baryon to baryon ({\alam}/{\lam}) ratio is considered as the measure of baryon transport number in $pp$ collisions to final state hadrons. It has been observed from the model comparison that none of the single models completely predict the experimental observations. However, QGSJET model predictions are close to the experimental results in the case of ratios as a function of rapidity $y$. HIJING, on the other hand, is in good agreement with {\alam}/{\ks} ratios as a function of {\ppt}. 

The {\alam}/{\ks} ratios as a function of {\ppt} shows the increasing trend at \sqrts~= 0.9, and 7 TeV suggest that in data, more baryons are expected to produce compared to mesons in strange hadronisation specifically at high {\ppt} regions. HIJING and QGSJET models support the experimental data and the above statement; however, Sibyll does not. Further improvement is required to fully understand the strange hadron production mechanisms, particularly at low {\ppt} regions. These ratios are further studied as a function of rapidity loss ($\Delta y$); all models fail to produce the experimental results. It is seen that models are somehow independent of rapidity. 

A current study of $V^0$ production ratios using various pQCD based models and cosmic-ray air shower based MC generators will be helpful for the development and hence, further improve the predictions of Standard Model physics at RHIC and LHC experiments, and it may also be beneficial in defining the baseline for discoveries. 

{\bf Data availability}
The data used to support the findings of this study are included within the article and are cited at relevant places within the
text as references.

{\bf Compliance with Ethical Standards}
The authors declare that they are in compliance with ethical standards regarding the content of this paper.
%

\begin{thebibliography}{99}

\bibitem{1}
H.~Yassin, E.~R.~A.~Elyazeed and A.~N.~Tawfik,
Phys. Scripta \textbf{95}, no.7, 7 (2020)
doi:10.1088/1402-4896/ab9128
[arXiv:1912.01404 [hep-ph]].

\bibitem{1a}
M. Waqas and G.-X. Peng and F.-H. Liu
 J. Phys. G: Nucl. Part. Phys. 48 075108 (2021)
doi:10.1088/1361-6471/abdd8d



\bibitem{1b}
M. Waqas, F.-H. Liu, R.-Q. Wang and I. Siddique
The European Physical Journal A 56, 188 (2020)
doi:10.1140/epja/s10050-020-00192-y


\bibitem{2}
M.~M.~Aggarwal \textit{et al.} [STAR],
Phys. Rev. C \textbf{83}, 024901 (2011)
doi:10.1103/PhysRevC.83.024901
[arXiv:1010.0142 [nucl-ex]].
\bibitem{3}
J.~Adam \textit{et al.} [ALICE],
Nature Phys. \textbf{13}, 535-539 (2017)
doi:10.1038/nphys4111
[arXiv:1606.07424 [nucl-ex]].



\bibitem{4}
J.~Adam \textit{et al.} [ALICE],
Phys. Lett. B \textbf{753}, 319-329 (2016)
doi:10.1016/j.physletb.2015.12.030
[arXiv:1509.08734 [nucl-ex]].

\bibitem{4a}
M. Ajaz , M. Waqas, L.-L. Li, A. Haj Ismail, U. Tabassam and M. Suleymanov
Eur. Phys. J. Plus \textbf{137}, 592 (2022)  https://doi.org/10.1140/epjp/s13360-022-02805-5

\bibitem{4b}
M. Ajaz, A. Haj Ismail, M. Waqas, M. Suleymanov, A. AbdelKader and R. Suleymanov
Scientific Reports \textbf{12}, 8142 (2022)  https://doi.org/10.1038/s41598-022-11685-9

\bibitem{1n}
V.~Khachatryan \textit{et al.} [CMS],
JHEP \textbf{09}, 091 (2010)
doi:10.1007/JHEP09(2010)091
[arXiv:1009.4122 [hep-ex]].

\bibitem{2n}
V.~Khachatryan \textit{et al.} [CMS],
Phys. Lett. B \textbf{765}, 193-220 (2017)
doi:10.1016/j.physletb.2016.12.009
[arXiv:1606.06198 [nucl-ex]].


\bibitem{3n}
J.~Adam \textit{et al.} [ALICE],
Nature Phys. \textbf{13}, 535-539 (2017)
doi:10.1038/nphys4111
[arXiv:1606.07424 [nucl-ex]].

\bibitem{ALICEn}
E.~Abbas \textit{et al.} [ALICE],
Eur. Phys. J. C \textbf{73}, 2496 (2013)
doi:10.1140/epjc/s10052-013-2496-5
[arXiv:1305.1562 [nucl-ex]].


\bibitem{10}
A.~E.~M.~Billmeier [STAR],
J. Phys. G \textbf{30}, S363-S368 (2004)
doi:10.1088/0954-3899/30/1/043
\bibitem{9}
D.~Colella [ALICE],
Int. J. Mod. Phys. Conf. Ser. \textbf{46}, 1860017 (2018)
doi:10.1142/S2010194518600170

\bibitem{LHCbn}
R.~Aaij \textit{et al.} [LHCb],
JHEP \textbf{08}, 034 (2011)
doi:10.1007/JHEP08(2011)034
[arXiv:1107.0882 [hep-ex]].

\bibitem{9a}
Muhammad Ajaz et al.,
Results in Physics \textbf{30}, 104790 (2021)
doi:10.1016/j.rinp.2021.104790

\bibitem{9b}
M. Ajaz, M. Waqas, G. X Peng et al. 
Eur. Phys. J. Plus \textbf{137}, 52 (2022). 
doi:10.1140/epjp/s13360-021-02271-5
[arXiv:2112.03187 [hep-ph]]

\bibitem{9c}
 S. Ullah, M. Ajaz, Z. Wazir, et al. 
 Sci Rep, \textbf{9}, 11811 (2019)
 doi:10.1038/s41598-019-48272-4

\bibitem{9d}
M. Ajaz, S. Ullah, Y. Ali, and H. Younis. 
Modern Physics Letters A, \textbf{33}, 1850038 (2018)
doi:10.1142/S0217732318500384


\bibitem{9e}
David d’Enterria, Ralph Engel, Tanguy Pierog, Sergey Ostapchenko, and Klaus Werner. 
Astroparticle Physics, \textbf{35} 98–113 (2011)
doi:10.1016/j.astropartphys.2011.05.002


\bibitem{9f}
S. Ullah, Y. Ali, M. Ajaz, U. Tabassam, and Q. Ali. 
International Journal of Modern Physics A, \textbf{33}, 1850108 (2018)
doi:10.1142/S0217751X18501087


\bibitem{9g}
  M. Ajaz, M. Tufail, and Y. Ali. 
  Arabian Journal for Science and Engineering, \textbf{45}, 411–416 (2019)

\bibitem{9h}
    S. Ullah, M. Ajaz, and Y. Ali. 
EPL, \textbf{123}, 31001 (2018)
doi:10.1209/0295-5075/123/31001

\bibitem{9i}
M. Ajaz, M. Bilal, Y. Ali, M. K. Suleymanov, and K. H. Khan.
Modern Physics Letters A, \textbf{34}, 1950090 (2019)
doi:10.1142/S0217732319500901

\bibitem{9j}
M. Ajaz, A.M. Khubrani, M. Waqas, A. Haj Ismail, E.A. Dawi,
Collective properties of hadrons in comparison of models prediction in pp collisions at 7 TeV,
Results in Physics, \textbf{36}, 105433 (2022)
doi:10.1016/j.rinp.2022.105433.

\bibitem{9k}
	P.-P. Yang, M. Ajaz, M. Waqas, F.-H. Liu, and M. Suleymanov. 
Journal of Physics G: Nuclear and Particle Physics, \textbf{49}, 055110 (2022)
doi:10.1088/1361-6471/ac5d0b

\bibitem{11}
M.H.M.  ~Soleiman, 
Arab J. Nucl. Sci. Appl., Vol. \textbf{53}, 1, 46-57 (2020)
doi: 10.21608/ajnsa.2019.9617.1182

\bibitem{12}
M.~Gyulassy and X.~N.~Wang,
Comput. Phys. Commun. \textbf{83}, 307 (1994)
doi:10.1016/0010-4655(94)90057-4
[arXiv:nucl-th/9502021 [nucl-th]].

\bibitem{13}
X.~N.~Wang and M.~Gyulassy,
Phys. Rev. D \textbf{45}, 844-856 (1992)
doi:10.1103/PhysRevD.45.844

\bibitem{14}
X.~N.~Wang and M.~Gyulassy,
Phys. Rev. D \textbf{44}, 3501-3516 (1991)
doi:10.1103/PhysRevD.44.3501

\bibitem{15}
D.~Kieda, M.~Salamon and B.~Dingus,
Proceedings, 26th International Cosmic Ray Conference (ICRC) : Contributed Papers: Salt Lake City, USA, August 17-25, 1999,

\bibitem{16}
F.~Riehn, R.~Engel, A.~Fedynitch, T.~K.~Gaisser and T.~Stanev,
Phys. Rev. D \textbf{102}, no.6, 063002 (2020)
doi:10.1103/PhysRevD.102.063002
[arXiv:1912.03300 [hep-ph]].

\bibitem{17}
F.~Riehn, H.~P.~Dembinski, R.~Engel, A.~Fedynitch, T.~K.~Gaisser and T.~Stanev,
PoS \textbf{ICRC2017}, 301 (2018)
doi:10.22323/1.301.0301
[arXiv:1709.07227 [hep-ph]].

\bibitem{18}
S.~Ostapchenko,
Nucl. Phys. B Proc. Suppl. \textbf{151}, 143-146 (2006)
doi: 10.1016/j.nuclphysbps.2005.07.026
[arXiv:hep-ph/0412332 [hep-ph]].


\bibitem{19}
S.~Ullah, M.~Ajaz and Y.~Ali,
EPL \textbf{123}, no.3, 31001 (2018)
doi: 10.1209/0295-5075/123/31001


\bibitem{20}
M. Ajaz et al., Modern Physics Letters A \textbf{34}, 1950090 (2019)
doi: 10.1142/S0217732319500901

\bibitem{21}
B.~I.~Abelev \textit{et al.} [STAR],
Phys. Rev. C \textbf{75}, 064901 (2007)
doi:10.1103/PhysRevC.75.064901
[arXiv:nucl-ex/0607033 [nucl-ex]].

\bibitem{22}
R.~Aaij \textit{et al.} [LHCb],
JHEP \textbf{08}, 034 (2011)
doi:10.1007/JHEP08(2011)034
[arXiv:1107.0882 [hep-ex]].



\bibitem{23}
X.~N.~Wang,
Phys. Rev. C \textbf{58}, 2321 (1998)
doi:10.1103/PhysRevC.58.2321
[arXiv:hep-ph/9804357 [hep-ph]].

\bibitem{24}
A.~D.~Martin, W.~J.~Stirling and R.~G.~Roberts,
Phys. Rev. D \textbf{50}, 6734-6752 (1994)
doi:10.1103/PhysRevD.50.6734
[arXiv:hep-ph/9406315 [hep-ph]].

\bibitem{25}
J.~Adams \textit{et al.} [STAR],
Phys. Lett. B \textbf{637}, 161-169 (2006)
doi:10.1016/j.physletb.2006.04.032
[arXiv:nucl-ex/0601033 [nucl-ex]].







  \end{thebibliography}
\end{document}